\documentclass[a4paper,11pt]{article}
\usepackage{t1enc}
\usepackage{amsmath}
\usepackage{amssymb}
\usepackage{enumerate}
\usepackage[latin9]{inputenc}
\usepackage[english]{babel}
\usepackage{fullpage}
\usepackage{verbatim}
\usepackage{psfrag}
\usepackage{braket}
\usepackage{color}
\usepackage{multicol,bbm}

\usepackage{jheppub}

\renewcommand{\eqref}[1]{\aref({#1})}

\newcommand{\vth}{\vartheta}

\newcommand{\Sm}[1]{\mathbf{S}_{#1}}
\newcommand{\Qm}[1]{\mathbf{Q}_{#1}}
\newcommand{\Um}[1]{\mathbf{U}_{#1}}
\newcommand{\Ima}[1]{\mathbf{I}_{#1}}
\newcommand{\Pm}[1]{\mathbf{P}_{#1}}
\newcommand{\Km}[1]{\mathbf{K}_{#1}}
\newcommand{\Rm}[1]{\mathbf{R}_{#1}}

\newcommand{\Mm}[1]{\mathbb{M}_{#1}}
\newcommand{\Mtm}[1]{\widetilde{\mathbb{M}}_{#1}}

\newcommand{\Am}[1]{\mathbb{A}_{#1}}
\newcommand{\Bm}[1]{\mathbb{B}_{#1}}
\newcommand{\Cm}[1]{\mathbb{C}_{#1}}
\newcommand{\Dm}[1]{\mathbb{D}_{#1}}
\newcommand{\Dvm}[1]{\overline{\mathbb{D}}_{#1}}
\newcommand{\Dhm}[1]{\widetilde{\mathbb{D}}_{#1}}
\newcommand{\Xm}[1]{\mathbb{X}_{#1}}
\newcommand{\X}[1]{\mathbf{X}_{#1}}

\newcommand{\Tm}[1]{\mathbb{T}_{#1}}

\newcommand{\Tmh}[2]{\widetilde{\mathbb{T}}_{#1}^{\Omega_{#2}}}

\newcommand{\Thm}[1]{\widehat{\mathbb{T}}_{#1}}
\newcommand{\hth}{\hat{\theta}}
\newcommand{\Res}[2]{\operatorname*{Res}_{#1=#2}}
\newcommand{\alphae}{\alpha^{\Omega_1}}
\newcommand{\betae}[1]{\beta_{#1}^{\Omega_1}}
\newcommand{\gammae}{\gamma^{\Omega_1}}
\newcommand{\deltae}[1]{\delta_{#1}^{\Omega_1}}
\newcommand{\taue}[1]{\tau^{\Omega_{#1}}}
\newcommand{\tauh}[1]{\widetilde{\tau}^{\Omega_{#1}}}
\newcommand{\Psie}[1]{\Ket{\Psi^{\Omega_{#1}}}}
\newcommand{\phie}{\Ket{\varphi^{\Omega_1}}_{\{i\}}}

\newcommand{\qe}[2]{q_{#1}^{(#2)}}

\newcommand{\Lam}{\Ket{\Lambda^{(0)}}}

\newcommand{\D}{\mathcal{D}}

\begin{document}

\title{Algebraic Bethe Ansatz for $O(2N)$ sigma models with integrable diagonal boundaries}

\author[a,b]{Tam\'as Gombor, }  
\author[b]{L\'aszl\'o Palla, }  
 \affiliation[a]{
MTA Lend\" ulet Holographic QFT Group, Wigner Research Centre,
H-1525 Budapest 114, P.O.B. 49, Hungary
} 
\affiliation[b]{
Institute for Theoretical Physics, Roland E\"otv\"os University,
1117 Budapest, P\'azm\'any s. 1/A Hungary}

{\hfill ITP-Budapest Report 667}

\emailAdd{gombor.tamas@wigner.mta.hu}
\emailAdd{palla@ludens.elte.hu}

\abstract{The finite volume problem of $O(2N)$ sigma models with integrable diagonal boundaries on a finite interval is investigated. The double row transfer 
matrix is diagonalized by Algebraic Bethe Ansatz. The boundary Bethe Yang equations for the particle rapidities and the accompanying Bethe Ansatz equations are derived.}

\maketitle

\section{Introduction}

Quantum integrable models in two dimensions 
have been the subject of intensive investigations recently, for at least two reasons. On the one hand they are interesting on their 
own, since in them one can compute many physically relevant quantities (e.g. 
finite volume spectra) exactly. On the other integrability based methods and 
specific integrable models play very important role in solving the 
spectral problem of AdS/CFT \cite{Beisert:2010jr}. 

A particularly interesting class of integrable models is the class of boundary 
integrable ones \cite{Ghoshal:1993tm}; 
they are defined on an open interval with integrability preserving boundary conditions at the ends of the interval\footnote{In these models the bulk interactions remain unchanged, the only change is that the particle like excitations have some additional - but integrability preserving - interactions with the boundaries.}. To solve these models one has to find first the reflection matrices (or boundary $S$ matrices) by solving the boundary Yang-Baxter, unitarity and crossing equations. The next step to solve their finite volume problem is 
to construct the double row monodromy (DM) and double row transfer (DTM)
matrices \cite{Sklyanin:1988yz}, since the boundary Bethe Yang equations for the particle rapidities are formulated with the help of the eigenvalues of the 
DTM. 

In this paper we consider the $O(2N)$ sigma model with integrability preserving diagonal boundaries.      
Sigma models with integrability preserving boundaries were considered both in the classical field theoretic framework \cite{MacKay:2001bh} \cite{MacKay:2004rz}
\cite{Mann:2006rh} \cite{Dekel:2011ja}, 
and in the quantum theory \cite{Ghoshal:1994bc}  
\cite{Moriconi:1998gc} \cite{Moriconi:2001xz}, where the various reflection 
matrices were obtained. We take the diagonal ones and construct with them the double row monodromy and transfer matrices. 

In the periodic case several methods have been devised to diagonalize the 
transfer matrices (and determine also the eigenvectors) of various 
 $D_N$ symmetric models \cite{Reshetikhin:1986vd} 
\cite{deVega:1986xj} \cite{Martins:1997wb}. We 
generalize the algebraic Bethe Ansatz (ABA) 
method of \cite{deVega:1986xj} to the boundary case to obtain the eigenvalues of our DTM.  

ABA has been used earlier to solve boundary problems. 
Building on earlier results \cite{deVega:1993xi} \cite{Mintchev:2001aq} 
\cite{Mezincescu:1990ui} the 
algebraic Bethe Ansatz was used to solve the $SU(N)$ vertex model with open 
boundaries \cite{Galleas:2004zj} while the open $GL(N)$ spin chain case was 
considered in \cite{Nepomechie:2009en}. In these cases the bulk $R$ matrix 
(which, apart from a scalar factor, coincides with the bulk $S$ matrix of the
sigma models) 
contains only two terms, one proportional to the identity and another one
proportional to the permutation matrices. In our case, as seen in
(\ref{eq:Skazakov}), there is a third term also, and its presence leads to 
difficulties already in the periodic case \cite{deVega:1986xj} .

The paper is organized as follows: in sect.2 we review the bulk 
$S$ matrix and the diagonal reflection matrices of the boundary $O(2N)$ sigma model and formulate the boundary Bethe Yang equations. In sect.3 we go through in details the diagonalization of our DTM 
when the reflection matrices at the two ends of the interval are identical, 
and building on these results we 
construct the Bethe Ansatz equations in sect.4. In sect.5 we summarize the same 
results for the case of different boundaries. 
We make our conclusions in 
sect.6. The paper is closed by three appendices, our notation is collected in 
Appendix A, while Appendix B and C are devoted to the derivation of 
some important technical details.      

\section{$O(2N)$ sigma model with integrable diagonal boundaries}


The quantum $O(2N)$ $\sigma$ model with integrable boundaries is specified by 
its $S$ matrix and reflection matrices describing the scattering of the
particles on each other and on the boundaries respectively. The particles 
transform according to the vector representation of $O(2N)$ and the simplest way
to describe the scattering processes is to use the 
Zamolodchikov - Fateev (ZF) 
operators \cite{Zamolodchikov:1980ku} associated to the particles $A_I^\dagger(\vth )$ where 
$I=1,\dots 2N$ and the 
rapidity $\vth $ determines the energy momentum of the particles as
\begin{equation}
p=m\sinh \left(\frac{\pi}{N-1}\vth\right),\qquad
E=m\cosh\left(\frac{\pi}{N-1}\vth\right) \,.
\end{equation}
The two particle $S$ matrix $S(\vth_1 ,\vth_2)$ and the reflection matrix 
$R(\vth)$ are encoded into the algebra of the ZF operators as
  \begin{equation}
 A_I^\dagger(\vth_1)A_J^\dagger(\vth_2)=
S_{IJ}^{KL}(\vth_1-\vth_2)A_L^\dagger(\vth_2)A_K^\dagger(\vth_1) ,\qquad
 A_I^\dagger(\vth)B=R_I^J(\vth)A_J^\dagger(-\vth )B,
\end{equation}
 where $B$ is the operator describing the boundary. As a result of the 
associativity of the ZF algebra the $S$ matrix and the reflection matrix 
satisfy the Yang Baxter (YB) and the boundary Yang Baxter (BYB) equations 
respectively:
\begin{equation}
  S_{I_1K_1}^{PQ}(\vth_2-\vth_3) S_{I_1Q}^{RK_2}(\vth_1-\vth_3)S_{RP}^{I_2J_2}(\vth_1-
\vth_2) =  S_{I_1J_1}^{PQ}(\vth_1-\vth_2) S_{PK_1}^{I_2R}(\vth_1-\vth_3)
S_{QR}^{J_2K_2}(\vth_2-\vth_3), \label{eq:YBE}
\end{equation}
\begin{equation}
  S_{I_1J_1}^{KL}(\vth_1-\vth_2) R_{K}^{P}(\vth_1) S_{LP}^{QI_2}(\vth_1+\vth_2) 
R_Q^{J_2}(\vth_2) =  R_{J_1}^L(\vth_2) S_{I_1L}^{KQ}(\vth_1+\vth_2) R_{K}^{P}(\vth_1) 
S_{QP}^{J_2I_2}(\vth_1-\vth_2). \label{eq:bYBE}
\end{equation}
In addition to these restrictions, on physical grounds, the scattering and 
reflection matrices satisfy the unitarity 
\begin{equation}
    S_{IJ}^{KL}(\vth_1-\vth_2)S_{LK}^{NM}(\vth_2-\vth_1)=\delta_I^M\delta_J^N,
\qquad  R_I^J(\vth)R_J^K(-\vth)=\delta_I^K,
   \end{equation}
crossing 
\begin{equation}
    S_{LI}^{JK}(i(N-1)-\vth)=S_{IJ}^{KL}(\vth), \label{eq:Scross}
   \end{equation}
and boundary crossing \cite{Ghoshal:1993tm}
 \begin{equation}
    S_{IK}^{LJ}(2\vth)R_{K}^{L}(i(N-1)-\vth)=R_I^J(\vth), \label{eq:Bcross}
   \end{equation}
conditions respectively.
 

For the $O(2N)$ sigma model the two particle $S$ matrix is known from the 
classic paper \cite{Zamolodchikov:1978xm}. Later on we separate its matrix and scalar parts thus we 
write it in the form\footnote{see Appendix A for our notation} 
\begin{equation}
\widehat{\Sm{AB}} =\sigma_2(\vth)\Sm{AB}=
\sigma_2(\vth)\Bigl[\Ima{AB}-\frac{i}{\vth}\Pm{AB} 
-\frac{i}{i(N-1)-\vth}\Km{AB} \Bigr], \label{eq:Skazakov}
\end{equation}
where
\begin{equation}
\Ima{AB}=\delta_{A_1}^{A_2}\delta_{B_1}^{B_2},\qquad
  \Pm{AB}=\delta_{A_1}^{B_2}\delta_{B_1}^{A_2},\qquad
  \Km{AB}=\delta_{A_1B_1}\delta^{A_2B_2}\ (=\Pm{AB}^{t_A}),
\end{equation}
and
\begin{equation}
\sigma_2(\vth)=\frac{\Gamma\Bigl(\frac{1}{2}+\frac{1}{2N-2}+\varphi\Bigr) 
			 \Gamma\Bigl(1+\varphi\Bigr)
			 \Gamma\Bigl(\frac{1}{2}-\varphi\Bigr)
			 \Gamma\Bigl(\frac{1}{2N-2}-\varphi\Bigr)}
			{\Gamma\Bigl(\frac{1}{2}+\frac{1}{2N-2}-\varphi\Bigr) 
			 \Gamma\Bigl(-\varphi\Bigr)
			 \Gamma\Bigl(\frac{1}{2}+\varphi\Bigr)
			 \Gamma\Bigl(1+\frac{1}{2N-2}+\varphi\Bigr)},\quad
\varphi=\frac{i\vth}{2N-2}.
\end{equation}
(Note that this matrix structure differs from the most well known $SU(N)$ 
case by the presence of the $\Km{AB}$ term). 

For the boundary case the solutions of (\ref{eq:bYBE}-\ref{eq:Bcross}) with 
 {\bf diagonal} $R$ matrices were found in \cite{Moriconi:2001xz}. 
The consistency of  
(\ref{eq:bYBE}) allows only two different elements in $R$
\begin{equation}
 R_A(\vth) = \mathrm{diag}(\underbrace{R_1,\dots,R_1}_{\text{K}},\underbrace{R_2,\dots,R_2}_{\text{2N-K}})=
R_2(\vth)\mathrm{diag}(\underbrace{c,\dots,c}_{\text K},\underbrace{1,\dots,1}_{\text 2N-K}) \label{eq:Rmat}
\end{equation}
with
\begin{equation}
c\equiv\frac{R_1(\vth)}{R_2(\vth)}=\frac{\frac{i}{2}(N-K)+\vth}{\frac{i}{2}(N-K)-\vth}.
\end{equation}
Physically these reflection matrices describe the case when $K$ of the 
fundamental fields satisfy Neumann, while the remaining $2N-K$ ones Dirichlet 
boundary conditions, (and generalize the pure Neumann ($K=2N$) and one Dirichlet
$2N-1$ Neumann cases investigated in \cite{Ghoshal:1994bc}). 
Note that these boundary conditions break the 
global $O(2N)$ symmetry of the bulk model to $O(K)\times O(2N-K)$. 
$R_2(\vth)$ is determined from the unitarity and boundary crossing conditions as
\begin{equation}
 R_2(\vth)=\frac{\Gamma\Bigl(\frac{1}{4}+\rho+\varphi\Bigr) 
		   \Gamma\Bigl(\frac{3}{4}+\rho-\varphi\Bigr)
		   \Gamma\Bigl(\frac{1}{4}+\rho (K-1)+\varphi\Bigr)
		   \Gamma\Bigl(\frac{3}{4}+\rho (K-1)-\varphi\Bigr)}
		  {\Gamma\Bigl(\frac{1}{4}+\rho-\varphi\Bigr) 
		   \Gamma\Bigl(\frac{3}{4}+\rho+\varphi\Bigr)
		   \Gamma\Bigl(\frac{1}{4}+\rho (K-1)-\varphi\Bigr)
		   \Gamma\Bigl(\frac{3}{4}+\rho (K-1)+\varphi\Bigr)}K(\vth),
\label{eq:r2expl}
\end{equation}
where $\rho=\frac{1}{4(N-1)}$ and $K(\vth)$ is the pure Neumann scalar factor
\begin{equation}
 K(\vth)=\frac{\Gamma\Bigl(\frac{1}{2}+\rho-\varphi\Bigr) 
		 \Gamma\Bigl(1+\varphi\Bigr)
		 \Gamma\Bigl(\frac{3}{4}+\rho+\varphi\Bigr)
		 \Gamma\Bigl(\frac{1}{4}-\varphi\Bigr)}
		{\Gamma\Bigl(\frac{1}{2}+\rho+\varphi\Bigr) 
		 \Gamma\Bigl(1-\varphi\Bigr)
		 \Gamma\Bigl(\frac{3}{4}+\rho-\varphi\Bigr)
		 \Gamma\Bigl(\frac{1}{4}+\varphi\Bigr)}.\label{eq:Neumann}
\end{equation}
 

\subsection{Boundary $O(2N)$ model in finite volume}

We are interested in the behaviour of a system of $n$ particles put into a 
strip of length $L$ with integrable boundaries at the ends of the strip 
(described by reflection matrices $R^{(+)}$, $R^{(-)}$ respectively) in the 
limit $L\rightarrow\infty$. (This condition guarantees that we may regard the 
system as a free gas with point like interactions).  The energy of the system 
is
\begin{equation}
 E \left(\{ \vth_i \}\right) = \sum_{i=1}^n m 
\cosh\left(\frac{\pi}{N-1}\vth_i\right) .
\end{equation}
The rapidities of the particles are determined by the boundary Bethe Yang 
equations
\begin{equation}
  e^{2ip(\vth_i)L}\prod_{j=i+1}^n\mathbb{S}_{i,j}(\vth_i-\vth_j)
\mathbb{R}_i^{(-)}(\vth_i)\prod_{j=n}^1\mathbb{S}_{j,i}(\vth_j+\vth_i)
\mathbb{R}_i^{(+)}(\vth_i)\prod_{j=1}^{i-1}\mathbb{S}_{i,j}(\vth_i-\vth_j)=
\mathbb{I},\ i=1,\dots n \label{eq:BBYE1}
 \end{equation}
where  $\mathbb{S}_{i,j}$ acts as the two particles $S$ matrix on the tensor 
product space of the i-th and j-th particles and as the identity on the 
others  (similarly $\mathbb{R}_i$ acts as $R$ on the i-th particles and as 
the identity on the others). This equation states that the total change in 
phase (including also the $\exp(i2p_iL)$ factor coming from free 
propagation)  vanishes for the process, in which the i-th particle is pushed 
through the others to the right boundary, gets reflected there, pushed through 
all the others until it gets to the left boundary, gets reflected, and again 
is pushed through until it gets back to its original position.

In the Bethe-Yang equation (\ref{eq:BBYE1}) the product of reflection and 
scattering matrices has to be diagonalized for all i. This can be done by 
introducing the double row monodromy and double row transfer 
matrices \cite{Sklyanin:1988yz}. 
To this end we introduce an auxiliary space $V^{2N}$ denoted by $A$ and define 
the double row monodromy matrix as
   \begin{equation}
  \omega_A(\vth;\{ \vth_i \})=\prod_{j=1}^n\mathbb{S}_{A,j}(\vth-\vth_j)
\mathbb{R}_A^{(-)}(\vth)\prod_{j=n}^1\mathbb{S}_{j,A}(\vth_j+\vth),
 \end{equation}
while the double row transfer matrix is given as
\begin{equation}
\mathcal{T}(\vth) = \mathrm{tr}_A\left[ \mathbb{R}_A^{(+) c}(i(N-1)-\vth) 
\omega_A(\theta) \right]. \label{eq:DTM}
\end{equation}
Here $\mathbb{R}^c$ denotes the charge conjugated reflection matrix 
$\mathbb{R}^c=\mathbb{C}\mathbb{R}\mathbb{C}^{-1}$. Please note that 
$\mathbb{R}^{(+) c}(i(N-1)-\vth)$ satisfies the boundary Yang Baxter, 
unitarity and boundary crossing conditions whenever 
$\mathbb{R}^{(+)}(\vth)$ does. Thus using the YB equation (\ref{eq:YBE}) and 
the BYB equation (\ref{eq:bYBE}) one can show that 
$[\mathcal{T}(\vth),\mathcal{T}(\lambda)]=0$, i.e. 
  $\mathcal{T}(\vth)$ and $\mathcal{T}(\lambda)$ can simultaneously be 
diagonalized. 
We define $\mathcal{T}$ this way \cite{Bajnok:2010ui}, because then 
(\ref{eq:BBYE1}) can be written as
\begin{equation}
e^{2ip(\vth)L}\mathcal{T}(\vth)\vert_{\vth=\vth_i}=-\mathbb{I} ,
\end{equation}
since according to (\ref{eq:Bcross}) $\mathbb{R}^{(+) c}((i(N-1)-\vth_i)$ and 
$\mathbb{S}(2\vth_i)$ combine into the required $\mathbb{R}^{(+)}(\vth_i)$. 
Thus, eventually, denoting by $\Lambda (\vth, \{\vth_i\})$ the eigenvalue of 
$\mathcal{T}(\vth)$ the boundary Bethe-Yang equations for the particle's 
rapidities can be written as
\begin{equation} 
e^{2ip(\vth_j)L}\Lambda(\vth_j,\{\vth_i\})=-1,\ j=1,\dots n \label{eq:BBYeq}
\end{equation}
In the bulk of the paper (in sect.3-4) 
we determine this eigenvalue when on the two ends of the strip the boundary conditions are identical $R^{(-)}(\vth)=R^{(+)}(\vth)$, and in sect.5 we summarize the results when the two reflection matrices are different.  

\section{Diagonalization of the DTM by algebraic BA}

We diagonalize the double row transfer matrix introduced in (\ref{eq:DTM}) 
by the algebraic Bethe Ansatz. 
To this end we adapt to the boundary case the procedure developed in 
\cite{deVega:1986xj} for the periodic case. 

\subsection{Definitions of the `reduced' monodromy and transfer matrices}
 

We find it convenient to introduce a new rapidity variable $\theta$ related to 
$\vth$ as
\[
\theta=\frac{2\vth}{i},\qquad \theta_j=\frac{2\vth_j}{i}
\]
and define 
\[
\hat{\theta}=2N-2-\theta=\frac{2}{i}(i(N-1)-\vth),
\]
such that the crossing transformation on the rapidity variable in 
(\ref{eq:Scross}) (\ref{eq:Bcross}) corresponds to 
$\theta\rightarrow\hat{\theta}$. Furthermore in the $2N$ dimensional target 
space $V^{2N}$ we introduce a complex basis
\[
\Ket{a}=\frac{1}{\sqrt{2}}(\Ket{2a-1}_r+i\Ket{2a}_r),\quad
 \Ket{\bar{a}}=\frac{1}{\sqrt{2}}(\Ket{2a-1}_r-i\Ket{2a}_r)\quad a=1,\dots,N.
 \]
corresponding to $V^{2N}=V^N+\bar{V}^N$. In this basis the matrix part of the 
two particle $S$ matrix in (\ref{eq:Skazakov}) takes the form
\begin{equation}
 \Sm{AB}=\begin{pmatrix}
			\Sm{ab} & 0 & 0 & 0\\
		  	0 & \Qm{ab} & \Um{ab} & 0\\
		  	0 & \Um{ab} & \Qm{ab} & 0\\
		  	0 & 0 & 0 & \Sm{ab}
	\end{pmatrix},
\end{equation}
where
\[
\Sm{ab}= \Ima{ab} - \frac{2}{\theta}\Pm{ab},\qquad
 \Qm{ab}= \Ima{ab} - \frac{2}{\hth}\Km{ab},\qquad
 \Um{ab}= - \frac{2}{\theta}\Pm{ab} - \frac{2}{\hth}\Km{ab}.
\]
When transforming also the reflection matrices (\ref{eq:Rmat}) into the 
complex basis it turns out that only for $K$ even, $K=2M$, is the reflection 
matrix also diagonal in this basis. Indeed in this case one finds 
\begin{equation}
R_A(\theta)=R_2(\theta)(\Rm{a}^0,\Rm{\bar{a}}^0)=R_2(\theta)\Rm{A} 
\label{eq:Rsep}
\end{equation} 
with
\[
\Rm{a}^0=\mathrm{diag}(\underbrace{c,\dots,c}_{\text M},\underbrace{1,\dots,1}_{\text N-M}),\quad c=\frac{N-2M+\theta}{N-2M-\theta},
\]
while for $K=2M-1$ $R(\theta)$ becomes non diagonal. Since we insist on having 
a diagonal reflection matrix we consider the $K=2M$ case only.
 
Since the diagonalization of the double row transfer matrix concerns 
the matrix parts we separate them from the 
scalar ones in eq.(\ref{eq:Skazakov}) (\ref{eq:Rsep})\footnote{Clearly the  
matrix parts alone satisfy the homogeneous Yang-Baxter and boundary Yang Baxter 
equations, and the matrix part $\Sm{AB}$ also satisfies 
$\Sm{AB}(\theta)=\Sm{AB}^{t_A}(\hth)$}. 
Thus we introduce the following \lq\lq reduced'' monodromy and transfer 
matrices
\begin{equation}
\mathcal{T}(\theta)=R_2(\theta)R_2(\hth)\prod\limits_{i=1}^n\sigma_2(\theta-\theta_i)\sigma_2(\theta+\theta_i)\D(\theta),
\end{equation}
\begin{equation}
\omega_A(\theta,\{\theta_i\})=R_2(\theta)\prod\limits_{i=1}^n\sigma_2(\theta-\theta_i)\sigma_2(\theta+\theta_i)\Mm{A}(\theta,\{\theta_i\}),
\end{equation}
 which are constructed from the matrix parts of eq.(\ref{eq:Skazakov}) 
(\ref{eq:Rsep}) only
\begin{align}
 \mathcal{D}(\theta;\{\theta_i\})&=\mathrm{tr}_A\bigl[\Rm{A}(\hth)\Mm{A}(\theta;\{\theta_i\})\bigr],\label{eq:rDTM}\\
 \Mm{A}(\theta;\{\theta_i\})&=\Tm{A}(\theta;\{\theta_i\})\Rm{A}(\theta)\Thm{A}(\theta;\{\theta_i\}),\label{eq4:mon}
\end{align}
where
\begin{equation}
 \Tm{A}(\theta;\{\theta_i\})=\Sm{An}(\theta-\theta_n) \cdots \Sm{A1}(\theta-\theta_1),\quad
 \Thm{A}(\theta;\{\theta_i\})=\Sm{1A}(\theta+\theta_1) \cdots \Sm{nA}(\theta+\theta_n).\label{eq:ThatT}
\end{equation}
Note that $\Thm{A}$ and $\Tm{A}$ depend on bulk quantities only (and $\Tm{A}$ 
is nothing but the periodic monodromy matrix),  
and using the 
crossing property one readily proves, that $\Thm{A}(\theta)=\Tm{A}^{t_A}(\hth)$.

After the repeated application of the bulk Yang-Baxter equation to change the order of the neighbouring $S$ 
matrices in the reduced double row transfer matrix in (\ref{eq:rDTM}-\ref{eq4:mon}) one can write
\[
\mathcal{D}(\theta,\{\theta_i\})=\mathcal{S}\mathrm{tr}_A\Bigl[\Rm{A}(\hth)\Sm{A1}(\theta-\theta_1)\dots
\Sm{An}(\theta-\theta_n)\Rm{A}(\theta)\Sm{nA}(\theta+\theta_n)\dots\Sm{1A}(\theta+\theta_1)\Bigr]\mathcal{S}^{-1} .
\]
Using the cyclic property of the trace the on the r.h.s. gives
\[
 \mathcal{D}(\theta;\{\theta_i\})=\mathcal{S}\mathrm{tr}_A\Bigl[ \Rm{A}(\theta)\Sm{An}^{t_n}(\theta+\theta_n) \cdots \Sm{A1}^{t_1}(\theta+\theta_1)
\Rm{A}(\hth)\Sm{1A}^{t_1}(\theta-\theta_1) \cdots \Sm{nA}^{t_n}(\theta-\theta_n)\Bigr]^{t}\mathcal{S}^{-1} ,
\]
where $t$ denotes transposition in the quantum space. Exploiting the crossing property of the bulk $S$ matrices this can be written as
\[
 \mathcal{D}(\theta;\{\theta_i\})=\mathcal{S}\mathrm{tr}_A\Bigl[ \Rm{A}(\theta)\Sm{An}(\hth-\theta_n) \cdots \Sm{A1}(\hth-\theta_1)
\Rm{A}(\hth)\Sm{1A}(\hth+\theta_1) \cdots \Sm{nA}(\hth+\theta_n)\Bigr]^{t}\mathcal{S}^{-1}. 
\]
Comparing to  (\ref{eq:rDTM}-\ref{eq4:mon}) leads to
\begin{equation}
\D(\theta ,\{\theta_i\})=\mathcal{S}\D^t(\hth,\{\theta_i\})\mathcal{S}^{-1},
\label{eq:dkrosszim}
\end{equation}
thus the eigenvalues of the reduced double row transfer matrix exhibit the crossing property 
\begin{equation}
\lambda(\theta ,\{\theta_i\})=\lambda(\hth,\{\theta_i\}).\label{eq:seszim}
\end{equation}

In the complex basis we write 
\[
 \Tm{A}(\theta)=\begin{pmatrix}
                 \alpha_a(\theta) & \beta_a(\theta)\\
                 \gamma_a(\theta) & \delta_a(\theta)
                \end{pmatrix},\quad
 \Thm{A}(\theta)=\begin{pmatrix}
                 \hat{\alpha}_a(\theta) & \hat{\beta}_a(\theta)\\
                 \hat{\gamma}_a(\theta) & \hat{\delta}_a(\theta)
                \end{pmatrix},    
 \]
while for the reduced double row monodromy matrix $\Mm{A}$
\[
 \Mm{A}(\theta)=\begin{pmatrix}
                 \Am{a}(\theta) & \Bm{a}(\theta)\\
                 \Cm{a}(\theta) & \Dm{a}(\theta)
                \end{pmatrix}.
\]
In this basis the double row transfer matrix - that we have to diagonalize - 
can be written as
\begin{equation}
\D(\theta)=\mathrm{tr}_a(\Rm{a}^0(\hth)\Am{a}(\theta))+
\mathrm{tr}_a(\Rm{a}^0(\hth)\Dm{a}(\theta)). \label{eq:DTMr}
\end{equation}

Implementing the algebraic Bethe Ansatz consists of three steps: first one has 
to find the 
so called pseudovacuum and the action of $\Am{a}$ and $\Dm{a}$ on it,  
then one has to determine the commutators among 
the elements of $\Mm{A}$, and finally (for $N\geq 2$) one has to find a 
recursion relation among the successive \lq\lq Bethe Ansatz steps'' - a 
procedure known as \lq\lq nesting''. We turn to these steps now.
 
 
\subsection{The pseudovacuum and the $\Am{a}(\theta)$, $\Dm{a}(\theta)$ operators}

To obtain the pseudovacuum and the action of the transfer matrix on it it is 
useful to write the elements of $\Mm{0}$ in terms of the elements of $\Tm{0}$ 
and $\Thm{0}$
\begin{align}
 \Am{a}(\theta)&=\alpha_a(\theta)\Rm{a}^0\hat{\alpha}_a(\theta)+
\beta_a(\theta)\Rm{a}^0\hat{\gamma}_a(\theta), \label{eq:Aexpl}\\
 \Bm{a}(\theta)&=\beta_a(\theta)\Rm{a}^0\hat{\delta}_a(\theta)+
\alpha_a(\theta)\Rm{a}^0\hat{\beta}_a(\theta),\\
 \Cm{a}(\theta)&=\gamma_a(\theta)\Rm{a}^0\hat{\alpha}_a(\theta)+
\delta_a(\theta)\Rm{a}^0\hat{\gamma}_a(\theta),\\
 \Dm{a}(\theta)&=\delta_a(\theta)\Rm{a}^0\hat{\delta}_a(\theta)+
\gamma_a(\theta)\Rm{a}^0\hat{\beta}_a(\theta). \label{eq:Dexpl}
\end{align}
We want $\Cm{a}$ to act as annihilation and $\Bm{a}$ as creation  operators 
on the pseudovacuum $\vert\phi\rangle$.   
Recalling the definitions of $\alpha_a,\dots ,\delta_a $ it is straightforward 
to see that one can achieve $\Cm{a}(\theta)\vert\phi\rangle =0$ if all the 
quantum particles belonging to 
$\vert\phi\rangle$ are in the unbarred space $V^N$, i.e. if
\begin{equation}
\vert\phi\rangle\in \Omega^0=V^N_n\otimes\cdot\cdot\cdot\otimes V_1^N.
\label{eq:omeganuldefi}
\end{equation}
On this pseudovacuum $\Bm{a}(\theta)$ indeed acts as a creation operator, but 
please note that it is creating a pair of bulk excitations. The second term in 
(\ref{eq:Aexpl}) annihilates any state in $\Omega^0$, thus on $\vert\phi\rangle$ 
$\Am{a}(\theta)$ may be replaced by
\begin{equation}
\Am{a}(\theta)\rightarrow 
\Sm{an}(\theta-\theta_n)\dots\Sm{a1}(\theta-\theta_1)\Rm{a}^0(\theta)\Sm{a1}(\theta+\theta_1)\dots\Sm{an}(\theta+\theta_n), \label{eq:Aact}
\end{equation}
i.e. $\Am{a}(\theta)$ has indeed a simple action. 
However we have to change the order of $\gamma_{a}(\theta)$ and 
$\hat{\beta}_{a}(\theta)$ in (\ref{eq:Dexpl}) to get a simple action for 
$\Dm{a}(\theta)$ on  $\vert\phi\rangle$. 

We compute this commutator starting from the bulk relation
\begin{equation}
\Tm{A}(\theta)\Sm{AB}(\theta+\theta^\prime)\Thm{B}(\theta^\prime)=
\Thm{B}(\theta^\prime)\Sm{AB}(\theta+\theta^\prime)\Tm{A}(\theta)
\label{eq:bulk1}\end{equation} 
which gives 
\[
\gamma_{a}(\theta)\Sm{ab}(\theta+\theta^\prime)\hat{\beta}_{b}(\theta^\prime)=
\hat{\beta}_{b}(\theta^\prime)\Sm{ab}(\theta+\theta^\prime)\gamma_{a}(\theta)+\\
\hat{\alpha}_{b}(\theta^\prime)\Um{ab}(\theta+\theta^\prime)\alpha_{a}(\theta)-
\delta_{a}(\theta)\Um{ab}(\theta+\theta^\prime)\hat{\delta}_{b}(\theta^\prime) .
\]
Since we are interested in the  action of $\Dm{a}(\theta)$ on 
 $\vert\phi\rangle$ we may neglect the first term, as it annihilates it. For 
the same reason, in all computations in the rest of this subsection we drop all 
terms ending with $\gamma$; we denote them simply by $\dots $. Using 
(\ref{eq:bulk1}) also to compute the terms arising from commuting 
$\hat{\alpha}_{b}(\theta^\prime)$ and $\alpha_{a}(\theta)$ together with the 
explicit forms and properties of $\Sm{ab}$, $\Um{ab}$ and $\Qm{ab}$ eventually 
we find
\begin{align}
\gamma_a(\theta)\Rm{a}^0\hat{\beta}_a(\theta)&=
\frac{2}{\theta-\hth}\Bigl( 
[\alpha_a(\theta)\Rm{a}^0\hat{\alpha}_a(\theta)]^{t_a}-
\mathrm{tr}_a[\alpha_a(\theta)\Rm{a}^0\hat{\alpha}_a(\theta)]-\nonumber\\
 &-\delta_{a}(\theta)[\Rm{a}^0]^{t_a}\hat{\delta}_{a}(\theta)+
\mathrm{tr}_a[\Rm{a}^0]\delta_{a}(\theta)\hat{\delta}_{a}(\theta)\Bigr)+\dots 
\label{eq:gammabetacom}
\end{align}
Since terms corresponding to $\Am{a}(\theta)$ appear here it is useful to 
introduce $\Dvm{a}(\theta)$ which is free from these terms
\begin{equation}
 \Dvm{a}(\theta)=\Dm{a}(\theta)+\frac{2}{\theta-\hth}\Bigl[\mathrm{tr}_a\Am{a}(\theta)-\Am{a}^{t_a}(\theta)\Bigr]\label{eq4:Dv}.
\end{equation}
Also combining the third and fourth terms in (\ref{eq:gammabetacom}) with 
(\ref{eq:Dexpl}) one finds
\[
\Dvm{a}(\theta)=\delta_{a}(\theta)\overline{\Rm{a}}^0\hat{\delta}_{a}(\theta)+\dots
\]
where we introduced the notation
\[
\overline{\Rm{a}}^0(\theta)=\Rm{a}^0(\theta)-\frac{2}{\theta-\hth}
\Bigl( [\Rm{a}^0(\theta)]^{t_a}-\mathrm{tr}_a[\Rm{a}^0(\theta)]\Bigr),
\]
which explicitly can be written as
\begin{multline}
\overline{\Rm{a}}^0(\theta)=\frac{\theta-N}{\theta+1-N}\cdot
\frac{\theta}{\theta+2M-N}\mathrm{diag}(\underbrace{-1,\dots ,-1}_{\text M},
\underbrace{1,\dots ,1}_{\text N-M}) 
=\\
\Bigl( \frac{\theta-N}{\theta+1-N}\cdot
\frac{\theta}{\theta+2M-N}\overline{\Rm{a}}^1(\theta)\Bigr) .
\label{eq:R1bar}
\end{multline}
(We defined $\overline{\Rm{a}}^1(\theta)$ for later reference).
Thus on any state in $\Omega^0$, including 
the pseudovacuum $\vert\phi\rangle$ $\Dvm{a}(\theta)$ may be replaced by
\begin{equation} 
\Dvm{a}(\theta)\rightarrow 
 \Qm{an}(\theta-\theta_n)\dots\Qm{a1}(\theta-\theta_1)
\overline{\Rm{a}}^0(\theta)
\Qm{a1}(\theta+\theta_1)\dots\Qm{an}(\theta+\theta_n).\label{eq:dbaract}
\end{equation}
Finally the transfer matrix (\ref{eq:DTMr}) takes the following form
\begin{equation}
\D(\theta)=\mathrm{tr}_a\bigl[\overline{\Rm{a}}^0(\hth)\Am{a}(\theta)\bigr]+
\mathrm{tr}_a\bigl[\Rm{a}^0(\hth)\Dvm{a}(\theta)\bigr] \label{eq:DTMr2}
\end{equation}
in terms of these quantities. 

Note that eq.(\ref{eq:omeganuldefi}) guarantees only that  $\Cm{a}(\theta)\vert\phi\rangle =0$, which is a neccesary condition
for any $\vert\phi\rangle$ being the pseudovacuum, i.e. the eigenvector of $\cal{D}(\theta)$. To obtain this eigenvector and the 
corresponding eigenvalue explicitly would require the diagonalization of (\ref{eq:DTMr2}) in $\Omega^0$ (a problem equivalent to solving 
an $SU(N)$ nesting problem). Instead of doing this now we proceed in deriving the general (i.e. non vacuum) eigenvalues of $\cal{D}(\theta)$ 
since for this (\ref{eq:Aact}) and (\ref{eq:dbaract}) are sufficient. From the expressions for the general eigenvalues the vacuum eigenvalue 
is obtained by specializing them to the absence of magnons.  

\subsection{The commutation relations between the elements of $\Mm{A}$}

In the framework of Algebraic Bethe Ansatz we look for eigenvectors of 
$\D(\theta)$ in the form
\begin{equation}
 \Ket{\Psi}=\Bigl[\prod_{r=1}^m\Bm{ }^{i_rj_r}(v_r)\Bigr]
\mathcal{F}_{i_1,..i_m,j_1,..j_m}\Ket{\phi} , \label{eq:pszi}
\end{equation}
(where $\mathcal{F}_{i_1,..i_m,j_1,..j_m}$ are rapidity independent constants) 
and determine under what conditions becomes $\ket{\Psi}$ an eigenvector
\footnote{Here we displayed the auxiliary space indexes of the $\Bm{ }$ 
operators explicitly. Also sometimes $\mathcal{F}_{i_1,..i_m,j_1,..j_m}\Ket{\phi}$ 
is abbreviated simply as $\Ket{\phi}_{\{i,j\}}$.}. To 
obtain these conditions one has to push $\Am{a}(\theta)$ and $\Dvm{a}(\theta)$ 
through the product of $\Bm{ }$-s, and to do this the commutation relations 
between these operators are needed.

We derive these commutation relations from the following equation satisfied by 
$\Mm{A}$ 
\begin{equation}
 \Sm{AB}(\theta-\theta')\Mm{A}(\theta)\Sm{AB}(\theta+\theta')\Mm{B}(\theta')=
\Mm{B}(\theta')\Sm{AB}(\theta+\theta')\Mm{A}(\theta)\Sm{AB}(\theta-\theta'). 
\label{eq4:YBE2}
\end{equation}  
This equation follows from the definition of $\Mm{a}(\theta)$ (\ref{eq4:mon}) and the fact that $\Sm{AB}(\theta)$  ($\Rm{A}^0(\theta)$) satisfy the Yang Baxter 
(\ref{eq:YBE}) (boundary Yang Baxter (\ref{eq:bYBE})) equations respectively 
\cite{Sklyanin:1988yz}
\footnote{Eq.(\ref{eq4:YBE2}) may be regarded as the generalization of the so called \lq\lq RTT'' relation to the boundary case}.

After some simple manipulations on the appropriate elements of 
eq.(\ref{eq4:YBE2}) one finds 
\begin{align}
 \Am{a}(\theta)&\Sm{ab}(\theta+u)\Bm{b}(u)=
 \Sm{ab}(u-\theta)\Bm{b}(u)\Qm{ab}(\theta+u)\Am{a}(\theta)\Qm{ab}^{-1}(u-\theta)-\nonumber\\
 &-\Bm{a}(\theta)\Qm{ab}(\theta+u)\Am{b}(u)\Um{ab}(u-\theta)\Qm{ab}^{-1}(u-\theta)-\Bm{a}(\theta)\Um{ab}(\theta+u)\Dm{b}(u).
\end{align}
To get a reasonable commutation relation we have to get rid of the $\Sm{ab}$ 
matrix standing between $\Am{a}(\theta)$ and $\Bm{b}(u)$ on the l.h.s. 
Therefore we take the transpose of this equation in the $b$ space, multiply 
both sides with the inverse of $\Sm{ab}^{t_b}(\theta +u)$, and take once more 
its transpose in the $b$ space to obtain
  \begin{align}
 \Am{a}(\theta)\Bm{b}(u)= &\biggl[\Bigl\{ \Sm{ab}(u-\theta)\Bm{b}(u) \Qm{ab}(u+\theta) \Am{a}(\theta) \Qm{ab}^{-1}(u-\theta) \Bigr\}^{t_b} \Qm{ab}^{-1}(\hat{u}-\theta) \biggr]^{t_b}-\nonumber\\
 -&\biggl[\Bigl\{ \Bm{a}(\theta) \Qm{ab}(\theta+u) \Am{b}(u) \Qm{ab}^{-1}(u-\theta)\Um{ab}(u-\theta) \Bigr\}^{t_b} \Qm{ab}^{-1}(\hat{u}-\theta) \biggr]^{t_b}-\nonumber\\
 -&\biggl[\Bigl\{\Bm{a}(\theta)\Um{ab}(\theta+u) \Dm{b}(u)\Bigr\}^{t_b}\Qm{ab}^{-1}(\hat{u}-\theta)\biggr]^{t_b}. \label{eq:abcom1}
\end{align}
The first term on the r.h.s. - where $\Am{a}$ and $\Bm{b}$ retained 
their argument - is the so called 'wanted term' 
(since looking for an eigenvector 
in the form of $\ket{\Psi}$ it may give a contribution), while the other two, 
that contain $\Bm{a}(\theta)$, are the 'unwanted ones' (as they can not 
contribute to an eigenvector in the form of $\ket{\Psi}$). It is important to 
establish a relation between the coefficients of the wanted and unwanted terms, 
as using this relation one can impose conditions on the wanted terms, 
guaranteeing the vanishing of the unwanted ones when constructing the 
eigenvectors. To this end, using the properties of the matrices $\Sm{ab}$,
$\Qm{ab}$ and $\Um{ab}$ summarized in \cite{deVega:1986xj}, 
we rewrite (\ref{eq:abcom1})
\begin{align}
 \Am{a}(\theta)\Bm{b}(u)= &\biggl[\Bigl\{ \Sm{ab}(u-\theta)\Bm{b}(u) \Qm{ab}(u+\theta) \Am{a}(\theta) \Qm{ab}^{-1}(u-\theta) \Bigr\}^{t_b} \Qm{ab}^{-1}(\hat{u}-\theta) \biggr]^{t_b}\nonumber\\
 -\frac{1}{\theta-u}\operatorname*{Res}_{\theta'=u}&\biggl[\Bigl\{ \Sm{ab}(u-\theta')\Bm{b}(\theta) \Qm{ab}(u+\theta) \Am{a}(\theta') \Qm{ab}^{-1}(u-\theta') \Bigr\}^{t_b} \Qm{ab}^{-1}(\hat{u}-\theta) \biggr]^{t_b}\nonumber\\
 +\frac{1}{\theta-\hat{u}}\operatorname*{Res}_{\theta'=u}&\biggl[\Sm{ab}(\theta'-u)\Bm{b}(\theta) \Dm{a}(\theta')^{t_a}\Qm{ab}^{-1}(\theta'-u)\biggr]^{t_b}
\label{eq4:CR_AB}.
\end{align}
In this form it is obvious, that the coefficient of the first unwanted term 
(i.e. the second term) is obtained as an appropriate residue of the 
coefficient of the wanted term -and in this respect eq.(\ref{eq4:CR_AB}) is 
similar to the analogous periodic expression in \cite{deVega:1986xj}. 
However, the 
third term  makes eq.(\ref{eq4:CR_AB}) significantly different from 
the analogous periodic expression: on the one hand this term has a different 
analytic structure then the first one and on the other it contains a new type 
of operator. Next 
we relate this term to the wanted term appearing in the 
$\Dvm{a}(\theta)\Bm{b}(u)$ commutator. 


First we rewrite eq.(\ref{eq4:CR_AB}) in terms of the $\Dvm{a}(\theta)$ 
operator introduced in eq.(\ref{eq4:Dv}). After some manipulations described 
in Appendix B we get
\begin{align}
 \Am{a}(\theta)\Bm{b}(u)= &\biggl[\Bigl\{ \Sm{ab}(u-\theta)\Bm{b}(u) \Qm{ab}(u+\theta) \Am{a}(\theta) \Qm{ab}^{-1}(u-\theta) \Bigr\}^{t_b} \Qm{ab}^{-1}(\hat{u}-\theta) \biggr]^{t_b}-\nonumber\\
 -\frac{1}{\theta-u}\operatorname*{Res}_{\theta'=u}&\biggl[\Bigl\{ \Sm{ab}(u-\theta')\Bm{b}(\theta) \Qm{ab}(u+\theta') \Am{a}(\theta') \Qm{ab}^{-1}(u-\theta') \Bigr\}^{t_b} \Qm{ab}^{-1}(\hat{u}-\theta') \biggr]^{t_b}-\nonumber\\
 +\frac{1}{\theta-\hat{u}}\operatorname*{Res}_{\theta'=u}&\biggl[\Sm{ab}(\theta'-u)\Bm{b}(\theta) \Dvm{a}(\theta')^{t_a}\Qm{ab}^{-1}(\theta'-u)\biggr]^{t_b}\label{eq4:comABDv}.
\end{align}       

We start to derive the $\Dvm{a}(\theta)\Bm{b}(u)$ commutator with the 
$\Dm{a}(\theta)\Bm{b}(u)$ one. From the appropriate elements of 
eq.(\ref{eq4:YBE2}) one obtains
\begin{align}
 \Dm{a}(\theta)\Qm{ab}(\theta+u)\Bm{b}(u)&=\Qm{ab}^{-1}(\theta-u)\Bm{b}(u)\Sm{ab}(\theta+u)\Dm{a}(\theta)\Sm{ab}(\theta-u)-\nonumber\\
 &-\Um{ab}(\theta-u)\Qm{ab}^{-1}(\theta-u)\Bm{a}(\theta)\Sm{ab}(\theta+u)\Dm{b}(u)+\nonumber\\
 &+\Qm{ab}^{-1}(\theta-u)\Am{b}(u)\Um{ab}(\theta+u)\Bm{a}(\theta)\Sm{ab}(\theta-u)-\nonumber\\
 &-\Um{ab}(\theta-u)\Qm{ab}^{-1}(\theta-u)\Am{a}(\theta)\Um{ab}(\theta+u)\Bm{b}(u).
\end{align}
We convert this into a reasonable commutator in the same way as in the previous 
case, i.e. by taking the transpose of the equation in $b$ space, multiplying 
both  sides by the inverse of $\Qm{ab}^{t_b}(\theta +u)$ and taking once more 
its transpose. However   
here, on the r.h.s., the order of $\Am{ }$-s and $\Bm{ }$-s is \lq wrong' as the 
latter ones stand on the right, thus we have to change them. When doing so, 
we get terms containing $\Bm{b}(u)\Am{a}(\theta)$, which is the \lq wanted' 
term in (\ref{eq4:comABDv}). We want to get rid of these terms as we want 
diagonal \lq wanted' terms in our commutation relations. We claim, that 
introducing  $\Dvm{a}(\theta)$, eq.(\ref{eq4:Dv}), in place of 
$\Dm{a}(\theta)$ does precisely this, and eventually in terms of this new 
operator we find
 \begin{align}
 \Dvm{a}(\theta)&\Bm{b}(u)=\biggl[\Bigl\{\Qm{ab}^{-1}(\theta-u)\Bm{b}(u)\Sm{ab}(\theta-\hat{u}+2)\Dvm{a}(\theta)\Sm{ab}(\theta-u) \Bigr\}^{t_b}\Sm{ab}(\theta-\hat{u})\biggr]^{t_b}-\nonumber\\
 &-\biggl[\Bigl\{\Um{ab}(\theta-u)\Qm{ab}^{-1}(\theta-u)\Bm{a}(\theta)\Sm{ab}(\theta-\hat{\theta}+2) \Dvm{b}(u)\Bigr\}^{t_b}\Sm{ab}(\theta-\hat{\theta})\biggr]^{t_b}+\nonumber\\
 &+\dots\label{eq4:com2}
\end{align}
where dots stand for the remaining unwanted terms containing only $\Am{b}(u)$ 
type operators.

What we really need is the commutation relations between the transfer 
matrix $\mathcal{D}(\theta)$, 
eq.(\ref{eq:DTMr2}), and $\mathbb{B}^{ij}(u)$. This problem is solved in Appendix C, where we establish the precise relation between the wanted and unwanted terms in these commutators. 
 

\subsection{First level of ABA}


The commutation relations presented in eqs.(\ref{eq4:comABDv}) (\ref{eq4:com2})
have a somewhat unusual form containing the transpositions. To emphasize that 
in spite of this they make it straightforward to evaluate the action of 
$\D(\theta)$ on the states generated by a chain of $\Bm{ }$-s from the 
pseudovacuum we consider first the action of $\Am{a}$ on a state obtained by 
using a single $\Bm{}$ only:
  \begin{align}
 &\Am{a}(\theta)\Bm{ }^{ij}(u)\mathcal{F}_{i,j}\Ket{\phi}=\nonumber\\
 &\biggl[\Bigl\{ \Sm{a1}(u-\theta)\Bm{1}(u) \Qm{a1}(u+\theta) \Am{a}(\theta) \Qm{a1}^{-1}(u-\theta) \Bigr\}^{t_1} \Qm{a1}^{-1}(\hat{u}-\theta) \biggr]^{ji}
\mathcal{F}_{i,j} \Ket{\phi}+\dots=\nonumber\\
 &\Bigl\{ \Sm{a1}(u-\theta)\Bm{1}(u) \Qm{a1}(u+\theta) \Am{a}(\theta) \Qm{a1}^{-1}(u-\theta) \Bigr\}^{kj} \Qm{ak}^{-1i}(\hat{u}-\theta)\mathcal{F}_{i,j} \Ket{\phi}
+\dots =\nonumber\\
 &\Sm{a}^{kl}(u-\theta)\Bm{l}^m(u) \Qm{am}^n(u+\theta) \Am{a}(\theta) \Qm{an}^{-1j}(u-\theta) \Qm{ak}^{-1i}(\hat{u}-\theta) \mathcal{F}_{i,j}\Ket{\phi}
+\dots =\nonumber\\
 &\Bm{l}^m(u)\Sm{a}^{t_1lk}(u-\theta) \Qm{am}^n(u+\theta) \Am{a}(\theta) \Qm{an}^{-1j}(u-\theta) \Qm{ak}^{-1i}(\hat{u}-\theta)\mathcal{F}_{i,j} \Ket{\phi}
+\dots =\nonumber\\
 &\Bm{ }^{lm}(u)\Bigl[\Sm{a1'}^{t_1}(u-\theta) \Qm{a1''}(u+\theta) \Am{a}(\theta) \Qm{a1''}^{-1}(u-\theta) \Qm{a1'}^{-1}(\hat{u}-\theta)\Bigr]_{lm}^{ij} 
\mathcal{F}_{i,j} \Ket{\phi}+\dots \nonumber
\end{align}      
where dots stand for the contribution of the \lq unwanted' terms. Using this 
the action of $\Am{a}(\theta)$ on $\ket{\Psi}$, (\ref{eq:pszi}), containing $m$ 
$\Bm{ }$ excitations is obtained as
 \begin{align}
 \Am{a}(\theta)\Ket{\Psi}=&\prod_{r=1}^m\Bm{ }^{k_rl_r}(v_r) \Bigl[\Sm{a1'}^{t_1}(v_1-\theta) \Qm{a1''}(v_1+\theta)\dots\Sm{am'}^{t_m}(v_m-\theta) \Qm{am''}(v_m+\theta)\times\nonumber\\ 
 \times&\Sm{an}(\theta-\theta_n)\dots\Sm{a1}(\theta-\theta_1)\Rm{a}^0(\theta)
\Sm{a1}(\theta+\theta_1)\dots\Sm{an}(\theta+\theta_n)\times\nonumber\\
 \times&\Qm{am''}^{-1}(v_m-\theta) \Qm{am'}^{-1}(\hat{v}_m-\theta)\dots \Qm{a1''}^{-1}(v_1-\theta) \Qm{a1'}^{-1}(\hat{v}_1-\theta)\Bigr]_{\{kl\}}^{\{ij\}} 
\Ket{\phi}_{\{i,j\}} +\dots ,
\end{align} 
where eq.(\ref{eq:Aact}) is used  on the r.h.s. to write the action of the 
$\Am{a}$ operator on the pseudovacuum, and 
dots denote the contribution of the \lq unwanted' terms. This shows that, similarly to the periodic case \cite{deVega:1986xj}, 
it is useful to enlarge the space $\Omega^0$ (\ref{eq:omeganuldefi}) to 
\begin{equation}
 \Omega_1 = (V_{1'}^N \otimes\cdots\otimes V_{m'}^N)\otimes(V_{n}^N \otimes\cdots\otimes V_{1}^N)\otimes(V_{m''}^N \otimes\cdots\otimes V_{1''}^N),
\end{equation}
and reinterpret the set of vectors $\Ket{\phi}_{\{i,j\}}$ as a vector in 
$\Omega_1$
\begin{equation}
  \Ket{\Psi^{\Omega_1}} \vert_{\Omega^0\in\Omega_1} = 
\Ket{\phi}_{\{i,j\}},\qquad
\Ket{\Psi^{\Omega_1}}\in\Omega_1 .
\end{equation}
We can rewrite the 
$\mathrm{tr}_a\bigl[\overline{\Rm{a}}^0(\hth)\Am{a}(\theta)\bigr]\ket{\Psi}$ 
part of $\D(\theta)\ket{\Psi}$ in terms of $\ket{\Psi^{\Omega_1}}$ by defining  
new monodromy and transfer matrices acting in $\Omega_1$. Indeed defining
\renewcommand{\Mm}[2]{\mathbb{M}_{#1}^{\Omega_#2}}
\renewcommand{\Mtm}[2]{\widetilde{\mathbb{M}}_{#1}^{\Omega_#2}}
\newcommand{\An}[2]{\mathbb{A}_{#1}^{\Omega_#2}}
\renewcommand{\Cm}[2]{\mathbb{C}_{#1}^{\Omega_#2}}
\renewcommand{\Dm}[2]{\mathbb{D}_{#1}^{\Omega_#2}}
\renewcommand{\Dhm}[1]{\widetilde{\mathbb{D}}_{#1}}
\newcommand{\Rvp}[2]{\overline{\mathbf{R}}_{#1}^{#2}}
\renewcommand{\Tm}[2]{\mathbb{T}_{#1}^{\Omega_{#2}}}
\renewcommand{\Tmh}[2]{\widehat{\mathbb{T}}_{#1}^{\Omega_{#2}}}
\renewcommand{\D}[1]{\mathcal{D}^{\Omega_{#1}}}
\begin{align}
 \Mm{a}{1}(\theta;\{\theta_i,v,w\}) =& \Sm{a1'}^{t_1}(v_1-\theta) \Sm{a1''}^{t_1}(\hat{w}_1-\theta) \dots \Sm{am'}^{t_m}(v_m-\theta) \Sm{am''}^{t_m}(\hat{w}_m-\theta) \times \nonumber \\ 
 \times & \Sm{an}(\theta-\theta_n) \dots \Sm{a1}(\theta-\theta_1) \Rm{a}^1(\theta) \Sm{a1}(\theta+\theta_1) \dots \Sm{an}(\theta+\theta_n) \times \nonumber \\
 \times & \Qm{am''}^{-1}(w_m-\theta) \Qm{am'}^{-1}(\hat{v}_m-\theta) \dots \Qm{a1''}^{-1}(w_1-\theta) \Qm{a1'}^{-1}(\hat{v}_1-\theta) , \\
 \D{1}(\theta;\{\theta_i,v,w\}) =& \mathrm{tr}_a \Bigl[ \Rvp{a}{1}(\hth) \Mm{a}{1}(\theta;\{\theta_i,v,w\}) \Bigr] \label{eq:D1def}
\end{align}
one finds
\begin{align}
\mathrm{tr}_a\bigl[&\overline{\Rm{a}}^0(\hth)\Am{a}(\theta)\bigr]\ket{\Psi^{\Omega_1}}=\nonumber\\
&\frac{\theta-N+2}{\theta-N+1}\cdot\frac{\theta-2N+2}{\theta-N-2M+2}
\prod\limits_{r=1}^{m}\Bm{ }^{i_rj_r}(v_r)\bigl[\D{1}(\theta;\{\theta_i,v,w\})
\Ket{\Psi^{\Omega_1}}\bigr]_{\{ij\}} +\dots ,\label{eq:nM1}
\end{align}
where dots stand for the \lq unwanted' terms, 
 we introduced $\Rm{a}^1(\theta)\equiv\Rm{a}^0(\theta)$, and also used 
$\overline{\Rm{a}}^1$ defined in (\ref{eq:R1bar}). For later convenience we also introduced the parameters $w_1,\dots ,w_m$; at the end of the procedure 
they have to be identified with $v_1,\dots ,v_m$ after a specific limit (
see \cite{deVega:1986xj}).

In an analogous way, defining
\newcommand{\Dh}[1]{\widetilde{\mathcal{D}}^{\Omega_{#1}}}
\newcommand{\Rp}[2]{\mathbf{R}_{#1}^{#2}}
\begin{align}
 \Mtm{a}{1}(\theta;\{\theta_i,v,w\}) =& \Qm{a1'}^{-1t_1}(\theta-v_1) \Qm{a1''}^{-1t_1}(\theta-\hat{w}_1) \dots \Qm{am'}^{-1t_m}(\theta-v_m) \Qm{am''}^{-1t_m}(\theta-\hat{w}_m) \times \nonumber \\ 
 \times & \Qm{an}(\theta-\theta_n) \dots \Qm{a1}(\theta-\theta_1) \Rvp{a}{1}(\theta) \Qm{a1}(\theta+\theta_1) \dots \Qm{an}(\theta+\theta_n) \times \nonumber \\
 \times & \Sm{am''}(\theta-w_m) \Sm{am'}(\theta-\hat{v}_m) \dots \Sm{a1''}(\theta-w_1) \Sm{a1'}(\theta-\hat{v}_1),\\
 \Dh{1}(\theta;\{\theta_i,v,w\}) =& \mathrm{tr}_a\Bigl[ \Rp{a}{1}(\hth) \Mtm{a}{1}(\theta;\{\theta_i,v,w\}) \Bigr] ,\label{eq:dtildefi}
\end{align}
the $\mathrm{tr}_a\bigl[\Rm{a}^0(\hth)\Dvm{a}(\theta)\bigr]\ket{\Psi}$ part of  
$\mathcal{D}(\theta)\ket{\Psi}$ can be written in terms of these quantities as
\begin{align}
\mathrm{tr}_a\bigl[&\Rm{a}^0(\hth)\Dvm{a}(\theta)\bigr]\ket{\Psi^{\Omega_1}}=
\nonumber \\
&\frac{\theta-N}{\theta-N+1}\cdot\frac{\theta}{\theta-N+2M}
\prod\limits_{r=1}^{m}\Bm{ }^{i_rj_r}(v_r)\bigl[\Dh{1}(\theta;\{\theta_i,v,w\}
)\Ket{\Psi^{\Omega_1}}\bigr]_{\{ij\}} +\dots \label{eq:nM2}
\end{align}

From the explicit form of $\Mm{a}{1}$ and $\Mtm{a}{1}$ and the fact that 
$\Rm{a}^1(\theta)$ satisfies the boundary Yang Baxter equation it follows that 
\begin{align}
  \Sm{ab}(\theta-\theta') \Mm{a}{1}(\theta) \Sm{ab}(\theta+\theta') \Mm{b}{1}(\theta') &= \Mm{b}{1}(\theta')\Sm{ab}(\theta+\theta')\Mm{a}{1}(\theta)\Sm{ab}(\theta-\theta'),\label{eq4:pYBE}\\
  \Qm{ab}(\theta-\theta') \Mtm{a}{1}(\theta) \Qm{ab}(\theta+\theta') \Mm{b}{1}(\theta') &= \Mm{b}{1}(\theta')\Qm{ab}(\theta+\theta')\Mtm{a}{1}(\theta)\Qm{ab}(\theta-\theta') 
\end{align}
hold. ($\Mtm{a}{1}$ also satisfies (\ref{eq4:pYBE})). Using these and the 
properties of $\Rm{a}^1$ one can prove, that $\D{1}(\theta)$ and 
$\Dh{1}(\theta^{\prime})$ commute
\[
[\D{1}(\theta),\D{1}(\theta^{\prime})]=
  [\D{1}(\theta),\Dh{1}(\theta^{\prime})]=
 [\Dh{1}(\theta),\Dh{1}(\theta^{\prime})]=0,
\]
i.e. they can be diagonalized simultaneously. 

On the basis of eq.(\ref{eq:nM1}) and (\ref{eq:nM2}) one can relate the 
eigenvalue $\lambda(\theta,\{\theta_i\})$ of the double row transfer matrix 
$\mathcal{D}(\theta)$ to the eigenvalues of $\D{1}(\theta)$ and 
$\Dh{1}(\theta)$ 
\newcommand{\Lamh}[1]{\widetilde{\Lambda}_{#1}}
\renewcommand{\Lam}[1]{\Lambda_{#1}}
 \begin{equation}
 \lambda = \frac{\theta-N+2}{\theta-N+1} \cdot \frac{\theta-2N+2}{\theta-N-2M+2} \Lam{1} + \frac{\theta-N}{\theta-N+1} \cdot \frac{\theta}{\theta-N+2M} \Lamh{1},\label{eq;rdtmsv}
\end{equation}
\begin{equation}
\D{1} \Psie{1} = \Lam{1} (\theta,\{\theta_i,v,w\}) \Psie{1},\qquad 
 \Dh{1} \Psie{1} = \Lamh{1} (\theta,\{\theta_i,v,w\}) \Psie{1},
\end{equation}
provided one can have a handle on the vanishing of the contributions of the unwanted terms (represented by dots) in (\ref{eq:nM1}) (\ref{eq:nM2}), i.e. if one 
can formulate the vanishing of these terms as some conditions on the eigenvalue 
$\lambda $. Since $\mathcal{D}(\theta)$ is independent of $v_i$ and $\ket{\Psi^{\Omega_1}}$ is independent of $\theta$ it is clear that
\begin{equation}
 \Res{\theta}{v_i} \lambda(\theta,\{\theta_i,v,v\}) = 0 \label{eq:resvanish}
\end{equation}
 is a necessary condition for this. On the basis of evidence we collected in Appendix C we argue that this condition is also sufficient.   

The $\D{1}(\theta)$ and $\Dh{1}(\theta)$ operators (together with their 
eigenvalues) can be related by an argument similar to the one employed to 
derive the crossing property of $\cal{D}(\theta)$, (\ref{eq:dkrosszim}). From 
the bulk YB equations one finds
\begin{align*}
 \Sm{ab}(\theta_a-\theta_b) \Sm{ac}(\theta_a-\theta_c) \Sm{bc}(\theta_b-\theta_c) &= \Sm{bc}(\theta_b-\theta_c) \Sm{ac}(\theta_a-\theta_c) \Sm{ab}(\theta_a-\theta_b), \\ 
 \Sm{ab}(\theta_a-\theta_b) \Qm{ac}(\theta_a-\theta_c) \Qm{bc}(\theta_b-\theta_c) &= \Qm{bc}(\theta_b-\theta_c) \Qm{ac}(\theta_a-\theta_c) \Sm{ab}(\theta_a-\theta_b),
\end{align*} 
while the crossing property of $\Sm{AB}$ implies 
$ \Qm{ab}(\theta) = \Sm{ab}^{t_a}(\hth)$. Using these equations and the argument
leading to (\ref{eq:dkrosszim}) one can show that $\D{1}$ and $\Dh{1}$ 
satisfy the following crossing relation
\[
 \Dh{1}(\theta,\{\theta_i,v_i,w_i\}) = \mathcal{O} \left[ \D{1} (\hth,\{\theta_i,v_i,w_i\}) \right]^{t_1}\mathcal{O}^{-1}, 
\]       
where $t_1$ denotes transposition in $\Omega_1$.                           
This implies that the eigenvalues satisfy
\begin{equation}
\Lamh{1} (\theta,\{\theta_i,v,w\})=\Lam{1} (\hth,\{\theta_i,v,w\}).
\label{eq:lamhlam}
\end{equation}
Please note that this also implies that the $\lambda(\theta,\{\theta_i\})$ 
eigenvalue of $\mathcal{D}(\theta)$, (\ref{eq;rdtmsv}), satisfies 
the crossing condition (\ref{eq:seszim}).

\subsection{Nesting}

In this subsection we determine the eigenvalue $\Lam{1}$ using the so called 
\lq nesting' procedure \cite{Babelon:1982gp} \cite{Kulish:1983rd} . 
The basic idea is the following: first we split the 
auxiliary space $V^N$ into the direct sum $V^N=V^1+V^{N-1}$ and write the 
monodromy matrix $\Mm{a}{1}$
\renewcommand{\Bm}[2]{\mathbb{B}_{#1}^{\Omega_#2}}
\renewcommand{\Dvm}[2]{\overline{\mathbb{D}}_{#1}^{\Omega_#2}}
\renewcommand{\Am}[1]{\mathbb{A}^{\Omega_#1}}
\begin{equation}
 \Mm{a}{1} = \begin{pmatrix}
 \Am{1} & \Bm{a}{1}\\
 \Cm{a}{1} & \Dm{a}{1}
 \end{pmatrix}, \label{eq:Malak}
\end{equation}
(where $\Am{1}$ is a scalar, $\Bm{a}{1}$ is an $N-1$ component row vector, 
$\Cm{a}{1}$ is a $N-1$ component column vector and $\Dm{a}{1}$ is a 
$(N-1)\times (N-1)$ matrix in the auxiliary space), and try to construct 
$\Lam{1}$ using a \lq\lq second'' Bethe Ansatz step. In this step we assume 
$\Bm{a}{1}$ as creation and $\Cm{a}{1}$ as annihilation operators using an 
appropriate pseudovacuum and express $\D{1}(\theta)$ in terms of $\Am{1}$ and 
$\Dm{a}{1}$. The term containing $\Am{1}$ has a simple action on 
$\ket{\Psi^{\Omega_1}}$, while the action of the term containing $\Dm{a}{1}$ can 
be described in terms of new monodromy and transfer matrices $\Mm{a}{2}$ and 
$\D{2}$ acting on a new space $\Omega_2$. Thus the eigenvalue problem of 
$\D{1}(\theta)$ is reduced to that of $\D{2}(\theta)$ and $\Lam{1}$ can be 
expressed as a known term (the action of $\Am{1}$) plus $\Lam{2}$ (the 
eigenvalue of $\D{2}$). In the next step we split $V^{N-1}=V^1+V^{N-2}$ and 
repeat the procedure. The success of these iterations depends on whether one 
is able to handle the various monodromy and transfer matrices appearing in the 
successive steps. If so then the procedure ends after the $N-2$-th step, when 
the new  monodromy matrix is a scalar in the auxiliary space. 

In the first \lq nesting' step the pseudovacuum is determined by the condition, that it is annihilated by 
$\Cm{a}{1}$ This condition is satisfied for the vectors 
\newcommand{\ui}[2]{u_{#1}^{(#2)}}
\renewcommand{\alphae}[1]{\alpha^{\Omega_#1}}
\newcommand{\alphe}[2]{\alpha_{#1}^{\Omega_#2}}
\renewcommand{\betae}[2]{\beta_{#1}^{\Omega_#2}}
\renewcommand{\gammae}[2]{\gamma_{#1}^{\Omega_#2}}
\renewcommand{\deltae}[2]{\delta_{#1}^{\Omega_#2}}
\renewcommand{\phie}[1]{\Ket{\varphi^{\Omega_{#1}}}_{\{i\}}}
\begin{equation}
 \phie{1} \in \Omega_1^{(0)} = \left( V_{1'}^{N-1} \otimes \cdots \otimes V_{m'}^{N-1}  \right) \otimes V_n \otimes \cdots \otimes V_1 \otimes\left( V_{m''}^{N-1} \otimes \cdots \otimes V_{1''}^{N-1}  \right)\label{eq:omega10def}
\end{equation}
thus we look for the eigenvectors of $\D{1}$ in the form
\begin{equation}
\Psie{1} = \prod_{r=1}^{n_1} \Bm{i_r}{1}(\ui{r}{1})\phie{1}\label{eq:psi1uj}    
\end{equation}

We need the actions of $\Am{1}$ and $\Dm{a}{1}$ on the pseudovacuum. To determine them it is useful 
to introduce the following operators
\begin{align}
 \Tm{a}{1}(\theta;\{\theta_i,v,w\}) =& \Sm{a1'}^{t_1}(v_1-\theta) \Sm{a1''}^{t_1}(\hat{w}_1-\theta)\dots\Sm{am'}^{t_m}(v_m-\theta) \Sm{am''}^{t_m}(\hat{w}_m-\theta)\times\nonumber\\ 
 &\times\Sm{an}(\theta-\theta_n)\dots\Sm{a1}(\theta-\theta_1)\\
 \Tmh{a}{1}(\theta;\{\theta_i,v,w\}) =&\Sm{a1}(\theta+\theta_1)\dots\Sm{an}(\theta+\theta_n)\times\nonumber\\
 &\times\Qm{am''}^{-1}(w_m-\theta) \Qm{am'}^{-1}(\hat{v}_m-\theta)\dots \Qm{a1''}^{-1}(w_1-\theta) \Qm{a1'}^{-1}(\hat{v}_1-\theta),
\end{align}
that depend on bulk quantities only. They satisfy the commutation relation
\begin{align}
  \Tm{a}{1}(\theta) \Sm{ab}(\theta+\theta') \Tmh{b}{1}(\theta') &= \Tmh{b}{1}(\theta')\Sm{ab}(\theta+\theta')\Tm{a}{1}(\theta).\label{eq:YBE2}
\end{align}
We write these new operators in the auxiliary space -similarly to $\Mm{a}{1}$  
- as 
\newcommand{\alphah}[1]{\hat{\alpha}^{\Omega_#1}}
\newcommand{\alphh}[2]{\hat{\alpha}_{#1}^{\Omega_#2}}
\newcommand{\betah}[2]{\hat{\beta}_{#1}^{\Omega_#2}}
\newcommand{\gammah}[2]{\hat{\gamma}_{#1}^{\Omega_#2}}
\newcommand{\deltah}[2]{\hat{\delta}_{#1}^{\Omega_#2}}
\begin{equation}
 \Tm{a}{1} = \begin{pmatrix}
 \alphae{1} & \betae{a}{1}\\
 \gammae{a}{1} & \deltae{a}{1}
 \end{pmatrix},\qquad\quad 
 \Tmh{a}{1} = \begin{pmatrix}
 \alphah{1} & \betah{a}{1}\\
 \gammah{a}{1} & \deltah{a}{1}
 \end{pmatrix},
\end{equation}
and using them one can write the $\Mm{a}{1}$, $\Am{1}$ and $\Dm{a}{1}$ 
operators as  
\renewcommand{\Rm}[2]{\mathbf{R}_{#1}^{#2}}
\begin{align}
 \Mm{a}{1} &= \Tm{a}{1} \Rm{a}{1} \Tmh{a}{1},\\
 \Am{1} &= \alphae{1} c_1 \alphah{1} + \betae{a}{1} \Rm{a}{1} \gammah{a}{1},\\
 \Dm{a_1a_2}{1} &= \deltae{a_1b_1}{1} \Rm{b_1b_2}{1} \deltah{ba_2}{1} + \gammae{a_1}{1} c_1 \betah{a_2}{1},\label{eq:DD1}
\end{align}
where (for later reference) we introduced $c_1(\theta)=c(\theta)$ in 
(\ref{eq:Rsep}). To compute the action of the $\Dm{a}{1}$ operator on the pseudovacuum we need the commutation relations of the $\gammae{a}{1}$, 
$\betah{a}{1}$ operators. One can find this from (\ref{eq:YBE2})
\begin{align}
 \gammae{a}{1}(\theta) \betah{b}{1}(\theta) = \frac{1}{\theta-1} \deltae{ac}{1}(\theta)\deltah{cb}{1}(\theta) - \frac{1}{\theta-1} \alphah{1}(\theta) \alphae{1}(\theta) + \frac{\theta}{\theta-1}\Sm{ac}^{db}(2\theta) \betah{c}{1}(\theta) \gammae{d}{1}(\theta),\nonumber
\end{align}
and substituting this into eq.(\ref{eq:DD1}) leads to
 \begin{equation}
 \Dm{a}{1}(\theta) = \deltae{a}{1}(\theta) \left[ \Rm{a}{1}(\theta) + \frac{c_1(\theta)}{\theta-1} \right] \deltah{a}{1}(\theta) - \frac{c_1(\theta)}{\theta-1} \alphah{1}(\theta) \alphae{1}(\theta) + \dots,\label{eq:DD2}
\end{equation}
where again dots represent the uninteresting 
terms ending with $\gamma$-s (that annihilate $\phie{1}$). Note that again a term corresponding to $\Am{1}$ appears here thus we change again $\Dm{a}{1}$ into 
$\Dvm{a}{1}$ which is free from this term
\begin{equation}
 \Dvm{a}{1}(\theta) = \Dm{a}{1}(\theta) + \frac{\Am{1}(\theta)}{\theta-1} .
\end{equation}
To express the $\D{1}(\theta)$ transfer matrix in terms of the new quantities 
we introduce 
\newcommand{\Rvm}[2]{\overline{\mathbf{R}}_{#1}^{#2}}
\[
\Rvm{a}{2}(\theta) = \mathrm{diag}(\underbrace{-1,\dots,-1}_{\text{M-1}},\underbrace{1,\dots,1}_{\text{N-M}}),
\]
appropriate for the $V^N=V^1+V^{N-1}$ decomposition, then from (\ref{eq:D1def}) 
we find 
 \begin{equation}
 \D{1}(\theta) = -\frac{\theta+N-2M}{\theta-1} \Am{1}(\theta) + \mathrm{tr}_a \left[ \Rvm{a}{2}(\hth) \Dvm{a}{1}(\theta) \right].\label{eq:D1redef}
\end{equation}

To obtain under what conditions (\ref{eq:psi1uj}) becomes an eigenvector of 
this $\D{1}(\theta)$ we need the commutation relations 
$\Am{1}(\theta)\Bm{a}{1}(u)$ and $\Dvm{a}{1}(\theta)\Bm{b}{1}(u)$. From 
(\ref{eq4:pYBE}) - using the explicit form of the various matrices - one can 
derive
\begin{align}
 \Am{1}(\theta)\Bm{a}{1}(u) =& \frac{\theta-u+2}{\theta-u} \frac{\theta+u}{\theta+u-2} \Bm{a}{1}(u)\Am{1}(\theta) - \frac{2}{\theta-u} \frac{\theta+u}{\theta+u-2} \Bm{a}{1}(\theta)\Am{1}(u) +\nonumber\\
 &+ \frac{2}{\theta+u-2} \Bm{a}{1}(\theta)\Dm{a}{1}(u),\label{eq:comA0}\\
 \Dm{a}{1}(\theta)\Bm{b}{1}(u) =& \Bm{b}{1}(u) \Sm{ab}(\theta+u) \Dm{a}{1}(\theta) \Sm{ab}(\theta-u) + \frac{2}{\theta-u} \Bm{b}{1}(\theta) \Sm{ab}(\theta+u) \Dm{a}{1}(u) \Pm{ab} -\nonumber\\
 &- \frac{2}{\theta-u} \frac{2}{\theta+u} \Am{1}(\theta)\Bm{b}{1}(u) - \frac{2}{\theta+u} \Am{1}(u)\Bm{b}{1}(\theta)\Pm{ab}\Sm{ab}(\theta-u) .\label{eq:comD0v}
\end{align}
Please note, that in (\ref{eq:comD0v}) the order of the $\Am{1}$ and 
$\Bm{b}{1}$ operators is wrong, changing them using (\ref{eq:comA0}) we get
  \begin{equation}
 \Dm{a}{1}(\theta)\Bm{b}{1}(u) = \Bm{b}{1}(u) \Sm{ab}(\theta+u-2) \Dm{a}{1}(\theta) \Sm{ab}(\theta-u) + \dots,\label{eq:DbBcom}
\end{equation}
where dots represent not only the unwanted terms but also some wanted ones 
containing $\Am{1}(\theta)$. We claim, that introducing $\Dvm{a}{1}$ in place 
of $\Dm{a}{1}$ just removes these terms and converts (\ref{eq:DbBcom}) to  
  \begin{equation}
 \Dvm{a}{1}(\theta)\Bm{b}{1}(u) = \Bm{b}{1}(u) \Sm{ab}(\theta+u-2) \Dvm{a}{1}(\theta) \Sm{ab}(\theta-u) + \dots ,\label{eq:DbBcom2}
\end{equation}
where $\dots$ represent now only the various unwanted terms. 

Using the commutation relations (\ref{eq:comA0}) and (\ref{eq:DbBcom2}) one can compute the action of $\D{1}(\theta)$, (\ref{eq:D1redef}), 
on $\ket{\Psi^{\Omega_1}}$. Indeed one finds
\begin{equation}
 \Am{1}(\theta) \Psie{1} = c_1(\theta) \prod_{i=1}^n \frac{\theta-\theta_i-2}{\theta-\theta_i} \cdot \frac{\theta+\theta_i-2}{\theta+\theta_i} \prod_{i=1}^{n_1} \frac{\theta-\ui{i}{1}+2}{\theta-\ui{i}{1}} \cdot \frac{\theta+\ui{i}{1}}{\theta+\ui{i}{1}-2} \Psie{1} + \dots,\label{eq:A1act}
\end{equation}
Computing the action of the $\mathrm{tr}_a \left[ \Rvm{a}{2}(\hth) \Dvm{a}{1}(\theta) \right]$ part of $\D{1}$ one realizes that it is useful to enlarge the 
space $\Omega_1^{(0)}$ (given in (\ref{eq:omega10def})) to $\Omega_2$
\begin{equation}
 \Omega_2 = \Omega_1^{(0)} \otimes \left( V_{n_1}^{N-1} \otimes \cdots \otimes V_1^{N-1} \right) ,
\end{equation}
introduce the \lq reduced' reflection matrix $\Rm{a}{2}(\theta)$ (obtained
from (\ref{eq:DD2}))
\[
\Rm{a}{2}(\theta) = \mathrm{diag}(\underbrace{c_2(\theta),\dots,c_2(\theta)}_{\text{M-1}},\underbrace{d_2(\theta),\dots,d_2(\theta)}_{\text{N-M}}),
\quad 
c_2(\theta) = \frac{\theta}{\theta-1} c(\theta), \quad
 d_2(\theta) = \frac{\theta-1+c(\theta)}{\theta-1} ,
\]
and define new monodromy and transfer matrices acting on $\Omega_2$:
\begin{align}
 \Mm{a}{2}(\theta) =& \Sm{a1}(\theta+\ui{1}{1}-2) \cdots  \Sm{an_1}(\theta+u_{n_1}^{(1)}-2) \times\nonumber\\
 &\Sm{a1'}^{t_{1'}}(v_1-\theta) \Sm{a1''}^{t_{1''}}(\hat{w}_1-\theta) \cdots \Sm{am'}^{t_{m'}}(v_m-\theta) \Sm{am''}^{t_{m''}}(\hat{w}_m-\theta) \Rm{a}{2}(\theta) \times \nonumber \\
 &\Qm{am''}^{-1}(w_m-\theta) \Qm{am'}^{-1}(\hat{v}_m-\theta) \cdots \Qm{a1''}^{-1}(w_1-\theta) \Qm{a1'}^{-1}(\hat{v}_1-\theta) \times\nonumber\\ 
&\Sm{an_1}(\theta-u_{n_1}^{(1)}) \cdots  \Sm{a1}(\theta-\ui{1}{1}),\\
 \D2(\theta) =& \mathrm{tr}_a \left[ \Rvm{a}{2}(\hth) \Mm{a}{2}(\theta) \right],
\end{align}
since using them one can write 
\begin{equation}
\mathrm{tr}_a \Rvm{a}{2}(\hth) \Dvm{a}{1}(\theta) \Psie{1} = \prod_{r=1}^{n_1}
\Bm{i_r}{1}(\ui{r}{1}) \left[ \D2(\theta) \Psie2 \right]_{\{i\}} + \dots,  
\quad\Psie2\in\Omega_2 .\label{eq:Db1act}
\end{equation}
(Here $\Psie2$ is an appropriate state in $\Omega_2$ 
\[
\Psie2\vert_{\Omega_1^{(0)}\in\Omega_2}= \phie{1} ;
\]
 its explicit form as a BA state is determined in the next nesting step).
This way the eigenvalue problem of $\D{1}(\theta)$ is reduced to that of 
$\D{2}(\theta)$ and the eigenvalue $\Lam{1}$ is related to $\Lam{2}$
\begin{equation}
 \Lam1 = \frac{\theta+N-2M}{\theta-1} \cdot \frac{\theta+N-2M}{\theta-N+2M} \prod_{i=1}^n \frac{\theta-\theta_i-2}{\theta-\theta_i}\cdot \frac{\theta+\theta_i-2}{\theta+\theta_i} \prod_{i=1}^{n_1} \frac{\theta-\ui{i}{1}+2}{\theta-\ui{i}{1}}\cdot \frac{\theta+\ui{i}{1}}{\theta+\ui{i}{1}-2} + \Lam2,
\end{equation}
where $ \D2 \Psie2 = \Lam2 (\theta,\{\theta_i,v,w,\ui{\ }{1}\}) \Psie2$. One
can prove, that the condition for the vanishing of the unwanted terms in 
(\ref{eq:A1act}) and (\ref{eq:Db1act}) is (see the end of App. C) 
\begin{equation}
 \Res{\theta}{\ui{j}{1}} \Lam1(\theta,\{ \theta_i,v,w,u^{(1)} \}) =0.
\end{equation}

In the $k$-th step we split $V^{N-(k-1)}=V^1+V^{N-k}$. In this step we repeat
the procedure we described in details for $k=1$, i.e. find the pseudovacuum,
\lq split' the monodromy matrix $\Mm{a}{k}$, find the 
modification $\Dvm{a}{k}$ having a simple
action on the pseudovacuum, find the commutation relations 
$\Am{k}(\theta)\Bm{a}{k}(u)$, $\Dvm{a}{k}(\theta)\Bm{b}{k}(u)$ and finally
find the action of $\D{k}(\theta)$ on the BA states generated by 
$\Bm{a}{k}(\ui{i}{k})$. We describe below only the essential points but do not
repeat all the details. 
 
The problem is that the monodromy and transfer matrices depend on whether 
$k\leq M$ or $M< k$. (Remember that $1\leq M\leq N$). To handle this
complication we define
\[
 c_{k+1}(\theta) = \frac{\theta}{\theta-k} c(\theta), \qquad
 d_{k+1}(\theta) = \frac{\theta-k+kc(\theta)}{\theta-k},
\]
and the new $R$ matrices (obtained from solving the iterations)
\begin{align}
 \Rm{a}{k+1}(\theta) &=
 \begin{cases}
  \mathrm{diag}(\overbrace{c_{k+1}(\theta),\dots,c_{k+1}(\theta)}^{\text{M-k}},\overbrace{d_{k+1},\dots,d_{k+1}}^{\text{N-M}}) & k=1,\dots,M , \\
  \mathrm{diag}(\underbrace{d_{k+1},\dots,d_{k+1}}_{\text{N-k}}) & k=M+1,\dots,N-1 , 
 \end{cases}  \\
 \Rvm{a}{k+1}(\theta) &= 
 \begin{cases}
  \mathrm{diag}(\overbrace{-1,\dots,-1}^{\text{M-k}},\overbrace{1,\dots,1}^{\text{N-M}}) & k=1,\dots,M , \\
  \mathrm{diag}(\underbrace{1,\dots,1}_{\text{N-k}}) & k=M+1,\dots,N-1 , 
 \end{cases}
\end{align}
(Note that not only $\mathbb{R}^2$, $\overline{\mathbb{R}}^2$ are of this
form, but also $\mathbb{R}^1$, $\overline{\mathbb{R}}^1$).

In the $k$-th step the monodromy and transfer matrices we start with have the
form 
 \begin{align}
 \Mm{a}{k}(\theta) =& \Sm{a1}(\theta+\ui{1}{k-1}-2(k-1)) \cdots  \Sm{an_{k-1}}(\theta+u_{n_{k-1}}^{(k-1)}-2(k-1)) \times\nonumber\\
 &\Sm{a1'}^{t_{1'}}(v_1-\theta) \Sm{a1''}^{t_{1''}}(\hat{w}_1-\theta) \cdots \Sm{am'}^{t_{m'}}(v_m-\theta) \Sm{am''}^{t_{m''}}(\hat{w}_m-\theta) \Rm{a}{k}(\theta) \times \nonumber \\
 &\Qm{am''}^{-1}(w_m-\theta) \Qm{am'}^{-1}(\hat{v}_m-\theta) \cdots \Qm{a1''}^{-1}(w_1-\theta) \Qm{a1'}^{-1}(\hat{v}_1-\theta) \times\nonumber\\ 
&\Sm{an_{k-1}}(\theta-u_{n_{k-1}}^{(k-1)}) \cdots  \Sm{a1}(\theta-\ui{1}{k-1}) , \label{eq:monk} \\
 \D{k}(\theta) =& \mathrm{tr}_a \left[ \Rvm{a}{k}(\hth) \Mm{a}{k}(\theta) \right] , \\
 \D{k}(\theta) \Psie{k} =& \Lam{k}(\theta) \Psie{k},\label{eq:ktheigenv}
\end{align}
satisfying the commutation relation
\begin{multline}
  \Sm{ab}(\theta-\theta') \Mm{a}{k}(\theta) \Sm{ab}(\theta+\theta'-2(k-1)) \Mm{b}{k}(\theta')= \\
  = \Mm{b}{k}(\theta')\Sm{ab}(\theta+\theta'-2(k-1))\Mm{a}{k}(\theta)\Sm{ab}(\theta-\theta').\label{eq4:pYBEk}
\end{multline}
Here, in (\ref{eq:ktheigenv}), $\Psie{k}$ is an appropriate vector in $\Omega_k$:
\[
\Omega_k=\Omega_{k-1}^{(0)}\otimes(V_1^{N-k+1}\otimes\dots\otimes V_{n_{k-1}}^{N-k+1}),
\]
where
\[
\Omega_{k-1}^{(0)}=(V_{1'}^{N-k+1}\otimes\dots\otimes V_{m'}^{N-k+1})\otimes(V_{m''}^{N-k+1}\otimes\dots\dots V_{1''}^{N-k+1})\otimes(V_1\otimes\dots\otimes V_{n_{k-2}}),
\]
which is constructed as a BA state. 
We write $\Mm{a}{k}$ in the auxiliary space in the form of (\ref{eq:Malak})
with $\Omega_2\rightarrow\Omega_k$ where $\Bm{a}{k}$ and $\Cm{a}{k}$ are $N-k$
component row and column vectors respectively. The appropriate operators 
\lq splitting' $\Mm{a}{k}$ can be written 
\begin{align}
 \Tm{a}{k}(\theta) =& \Sm{a1}(\theta+\ui{1}{k-1}-2(k-1)) \cdots  \Sm{an_{k-1}}(\theta+u_{n_{k-1}}^{(k-1)}-2(k-1)) \times\nonumber\\
 & \Sm{a1'}^{t_1}(v_1-\theta) \Sm{a1''}^{t_1}(\hat{w}_1-\theta)\dots\Sm{am'}^{t_m}(v_m-\theta) \Sm{am''}^{t_m}(\hat{w}_m-\theta)\\ 
 \Tmh{a}{k}(\theta) =& \Qm{am''}^{-1}(w_m-\theta) \Qm{am'}^{-1}(\hat{v}_m-\theta)\dots \Qm{a1''}^{-1}(w_1-\theta) \Qm{a1'}^{-1}(\hat{v}_1-\theta)\times\nonumber\\
&\Sm{an_{k-1}}(\theta-u_{n_{k-1}}^{(k-1)}) \cdots  \Sm{a1}(\theta-\ui{1}{k-1}),
\end{align}
with commutation relations
\begin{align}
  \Tm{a}{k}(\theta) \Sm{ab}(\theta+\theta'-2(k-1)) \Tmh{b}{k}(\theta') &= \Tmh{b}{k}(\theta')\Sm{ab}(\theta+\theta'-2(k-1))\Tm{a}{k}(\theta).\label{eq:YBE2k}
\end{align}
In terms of these operators the splitting can be described as 
\begin{align}
 \Mm{a}{k} &= \Tm{a}{k} \Rm{a}{k} \Tmh{a}{k},\\
 \Am{k} &= \alphae{k} c_k \alphah{k} + \betae{a}{k} \Rm{a}{k} \gammah{a}{k},\\
 \Dm{a_1a_2}{k} &= \deltae{a_1b}{k} \Rm{a}{k} \deltah{ba_2}{k} + \gammae{a_1}{k} c_k \betah{a_2}{k}.\label{eq:DDk}
\end{align}
From the commutation relations (\ref{eq:YBE2k}) it follows that
\begin{multline}
 \gammae{a}{k}(\theta) \betah{b}{k}(\theta) = \frac{1}{\theta-k} \deltae{ac}{k}(\theta)\deltah{cb}{k}(\theta) - \frac{1}{\theta-k} \alphah{k}(\theta) \alphae{k}(\theta) +\\
 + \frac{\theta}{\theta-k}\Sm{ac}^{db}(2\theta-2(k-1)) \betah{c}{k}(\theta) \gammae{d}{k}(\theta).
\end{multline}
Using this in (\ref{eq:DDk}) one finds
\begin{equation}
 \Dm{a}{k}(\theta) = \deltae{a}{k}(\theta) \left[ \Rm{a}{k}(\theta) + \frac{c_k(\theta)}{\theta-k} \right] \deltah{a}{k}(\theta) - \frac{c_k(\theta)}{\theta-k} \alphah{k}(\theta) \alphae{k}(\theta) + \dots
\end{equation}
In this case one can remove the terms corresponding to $\Am{k}$ by introducing
$\Dvm{a}{k}$
\begin{equation}
 \Dvm{a}{k}(\theta) = \Dm{a}{k}(\theta) + \frac{\Am{k}(\theta)}{\theta-k}  ,
\end{equation}
and finally $\D{k}$ can be written as
\begin{equation}
 \D{k}(\theta) =
 \begin{cases}
  -\frac{\theta+N-2M}{\theta-k} \Am{k}(\theta) + \mathrm{tr}_a \left[ 
\Rvm{a}{k+1}(\hth) \Dvm{a}{k}(\theta) \right] & k=1,\dots,M \\
  \frac{\theta-N}{\theta-k} \Am{k}(\theta) + \mathrm{tr}_a \left[ \Rvm{a}{k+1}(\hth) \Dvm{a}{k}(\theta) \right] & k=M+1,\dots,N-1 
 \end{cases}\label{eq:kikD}
\end{equation}
To compute the action of $\D{k}$ on the BA states  $\Psie{k}$ 
 generated by a chain of 
$\Bm{a}{k}$-s (appropriately generalizing (\ref{eq:psi1uj})), we need the commutators $\Am{k}(\theta)\Bm{a}{k}(u)$ and $\Dvm{a}{k}(\theta)\Bm{b}{k}(u)$, which can be obtained from  
(\ref{eq4:pYBEk}) (taking into account also the transition from $\Dm{a}{k}$ to
$\Dvm{a}{k}$):
\begin{align}
 \Am{k}(\theta)\Bm{a}{k}(u) =& \frac{\theta-u+2}{\theta-u}\cdot \frac{\theta+u-2(k-1)}{\theta+u-2k} \Bm{a}{k}(u)\Am{k}(\theta) + \dots \\
 \Dvm{a}{k}(\theta)\Bm{b}{k}(u) =& \Bm{b}{k}(u) \Sm{ab}(\theta+u-2k) \Dvm{a}{k}(\theta) \Sm{ab}(\theta-u) + \dots .
\end{align}
Repeating the steps we made for $k=1$ we find using these commutators that
\begin{align}
 \Am{k}(\theta) \Psie{k} =& c_k(\theta) \prod_{i=1}^{n_{k-1}} \frac{\theta-\ui{i}{k-1}-2}{\theta-\ui{i}{k-1}}\cdot \frac{\theta+\ui{i}{k-1}-2k}{\theta+\ui{i}{k-1}-2(k-1)}\times\nonumber\\
 & \prod_{i=1}^{n_k} \frac{\theta-\ui{i}{k}+2}{\theta-\ui{i}{k}}\cdot \frac{\theta+\ui{i}{k}-2(k-1)}{\theta+\ui{i}{k}-2k} \Psie{k} + \dots, \label{eq:Akact}\\
 \mathrm{tr}_a\Rvm{a}{k+1}(\hth) \Dvm{a}{k}(\theta) \Psie{k} =& \prod_{r=1}^{n_k} \Bm{i_r}{k}(\ui{r}{k}) \left[ \D{k+1}(\theta) \Psie{k+1} \right]_{\{i\}} + \dots, 
\label{eq:Dbkact}
\end{align}
where $\D{k+1}$ (and the monodromy matrix belonging to it) are given by 
(\ref{eq:monk}). Note that (\ref{eq:A1act}) is the form of (\ref{eq:Akact}) if 
one identifies 
\begin{equation}
\theta_i\equiv \ui{i}{0}\qquad {\rm and} \qquad n_0=n. \label{eq:uinul} 
\end{equation}
According to the end of App. C the condition 
for the vanishing of the unwanted terms (denoted by the dots in (\ref{eq:Akact}) (\ref{eq:Dbkact})) is
\begin{equation}
 \Res{\theta}{\ui{j}{k}} \Lam{k}(\theta,\{ \theta_i,v,w,u^{(k)} \}) =0.
\label{eq:kikLam}
\end{equation}

Finally, summing up all the terms (and using the notation in (\ref{eq:uinul})),
 one finds the eigenvalue $\Lam{1}$ in the form
\begin{equation}
 \Lam1 = \sum_{k=0}^{N-1} F_k,\label{eq:Lam1vegs}
\end{equation}
where
\begin{multline}
 F_k =  \frac{\theta+N-2M}{\theta-k-1} \cdot \frac{\theta}{\theta-k} \cdot \frac{\theta+N-2M}{\theta-N+2M}  \prod_{i=1}^{n_{k}} \frac{\theta-\ui{i}{k}-2}{\theta-\ui{i}{k}}\cdot \frac{\theta+\ui{i}{k}-2k-2}{\theta+\ui{i}{k}-2k} \times \\
  \prod_{i=1}^{n_{k+1}} \frac{\theta-\ui{i}{k+1}+2}{\theta-\ui{i}{k+1}}\cdot \frac{\theta+\ui{i}{k+1}-2k}{\theta+\ui{i}{k+1}-2k-2} \quad k=0,\dots,M-1, 
\end{multline}
\begin{multline}
 F_k =  \frac{\theta-N}{\theta-k-1} \cdot \frac{\theta}{\theta-k} \cdot \frac{\theta-N}{\theta-N+2M}  \prod_{i=1}^{n_{k}} \frac{\theta-\ui{i}{k}-2}{\theta-\ui{i}{k}}\cdot \frac{\theta+\ui{i}{k}-2k-2}{\theta+\ui{i}{k}-2k} \times \\
  \prod_{i=1}^{n_{k+1}} \frac{\theta-\ui{i}{k+1}+2}{\theta-\ui{i}{k+1}}\cdot \frac{\theta+\ui{i}{k+1}-2k}{\theta+\ui{i}{k+1}-2k-2} \quad k=M,\dots,N-2, 
\end{multline}
\begin{multline}
 F_{N-1} =  \frac{\theta}{\theta-N+1} \cdot \frac{\theta-N}{\theta-N+2M} \prod_{i=1}^{n_{N-1}} \frac{\theta-\ui{i}{N-1}-2}{\theta-\ui{i}{N-1}}\cdot \frac{\theta-\hat{u}_i^{(N-1)}-2}{\theta-\hat{u}_i^{(N-1)}}\times\\
  \prod_{i=1}^{m} \frac{\theta-v_i+2}{\theta-v_i}\cdot \frac{\theta-\hat{v}_i}{\theta-\hat{v}_i-2}\cdot \frac{\theta-w_i}{\theta-w_i-2}\cdot \frac{\theta-\hat{w}_i+2}{\theta-\hat{w}_i} . 
\end{multline}
The cancellation of the unwanted terms in the various steps is guaranteed by
\[
\frac{F_{k-1}}{F_k}\vert_{\theta=\ui{i}{k}}=-1
\]
\newcommand{\mi}[2]{\mu_{#1}^{(#2)}}

The nesting procedure discussed here can be used also to determine the 
$\Lamh{1} (\theta,\{\theta_i,v,w\})$ eigenvalue of 
$\Dh{1}(\theta;\{\theta_i,v,w\})$ (\ref{eq:dtildefi}). We claim, that the 
result can indeed be summarized by equation (\ref{eq:lamhlam}). 

\section{The Bethe Ansatz equations}

The eigenvalue of the reduced transfer matrix $\mathcal{D}(\theta)$ is obtained 
from (\ref{eq;rdtmsv}), (\ref{eq:Lam1vegs}) and (\ref{eq:lamhlam}). Starting 
from here one can 
construct the Bethe Ansatz equations in the same way as in the periodic 
case \cite{deVega:1986xj}, i.e. one has 
to let $w_i\rightarrow v_i$ in a certain specific way. In this limit the cancellation of certain poles at $\theta=w_i+2$ is achieved only if each $v_i$ coincides with one $\ui{i}{N-1}$. To describe the outcome it is useful to introduce a new notation, where the set of $\ui{i}{N-1}$ is divided into two sets 
 \begin{align}
 \ui{i}{N-1} &= \frac{2\mi{i}{+}}{i}+N-1,\quad & i&=1,\dots,n_+=m,\\
 \ui{i+m}{N-1} &= \frac{2\mi{i}{-}}{i}+N-1,\quad & i&=1,\dots,n_-=n_{N-1}-m,\\
 \theta_i &\equiv \ui{i}{0}=\frac{2\vth_i}{i}\equiv\frac{2\mi{i}{0}}{i},\quad & i&=1,\dots,n_0=n,\\
 \ui{i}{k} &= \frac{2\mi{i}{k}}{i}+k,\quad & i&=1,\dots,n_k,\ k=1,\dots,N-2.
\end{align}
 (Here we also made some shifts in the magnon rapidities making possible to 
cast the resulting  BAE into a more familiar form). Returning also to the rapidity variable $\vth$, 
($\theta=\frac{2\vth}{i}$), used in section 2, the eigenvalue of the reduced transfer matrix can be written as
\begin{align}
\lambda(\vth ,\{\mi{i}{k}\})&=
\frac{\vth -\frac{i}{2}(N-2)}{\vth-\frac{i}{2}(N-1)}\cdot
\frac{\vth -i(N-1)}{\vth-\frac{i}{2}(N+2M-2)}\sum\limits_{k=0}^{N-1}G_k(\vth) 
\nonumber\\ &+
\frac{\vth -\frac{i}{2}N}{\vth-\frac{i}{2}(N-1)}\cdot
\frac{\vth}{\vth-\frac{i}{2}(N-2M)}\sum\limits_{k=0}^{N-1}\tilde{G}_k(\vth).
\label{eq:kislam}\end{align}
To express $G_k(\vth)$ in a compact form we define
\[
Z_k(\vth)=\prod\limits_{i=1}^{n_k}
\frac{\vth-\mi{i}{k}+\frac{i}{2}}{\vth-\mi{i}{k}-\frac{i}{2}}\cdot
\frac{\vth+\mi{i}{k}+\frac{i}{2}}{\vth+\mi{i}{k}-\frac{i}{2}},\qquad
k=0,\dots,N-2,+,-,
\]
since using them one can write
\begin{align}
G_k(\vth)=\frac{\vth(\vth+\frac{i}{2}(N-2M))^2}{(\vth-\frac{i}{2}k)(\vth-\frac{i}{2}(k+1))(\vth-\frac{i}{2}(N-2M))}& Z_k^{-1}(\vth-\frac{i}{2}(k+1))
Z_{k+1}(\vth-\frac{i}{2}k),\label{eq:gk1}\\ &k=0,\dots,M-1\nonumber\\
G_k(\vth)=\frac{\vth(\vth-\frac{i}{2}N)^2}{(\vth-\frac{i}{2}k)(\vth-\frac{i}{2}(k+1))(\vth-\frac{i}{2}(N-2M))}& Z_k^{-1}(\vth-\frac{i}{2}(k+1))
Z_{k+1}(\vth-\frac{i}{2}k),\label{eq:gk2}\\ &k=M,\dots,N-3\nonumber \\
G_{N-2}(\vth)=\frac{\vth(\vth-\frac{i}{2}N)^2}{(\vth-\frac{i}{2}(N-1))(\vth-\frac{i}{2}(N-2))(\vth-\frac{i}{2}(N-2M))}& Z_{N-2}^{-1}(\vth-\frac{i}{2}(N-1))\times
\nonumber\\
Z_{-}(\vth-\frac{i}{2}(N-2))& Z_{+}(\vth-\frac{i}{2}(N-2)),\label{eq:gn2}\\
G_{N-1}(\vth)=\frac{\vth(\vth-\frac{i}{2}N)}{(\vth-\frac{i}{2}(N-1))(\vth-\frac{i}{2}(N-2M))}& Z_{-}^{-1}(\vth-\frac{i}{2}N)Z_{+}(\vth -\frac{i}{2}(N-2)) \label{eq:gnnm1},
\end{align}
and 
\[
\tilde{G}_{k}(\vth)=G_k(\hat{\vth}),\qquad \hat{\vth}=i(N-1)-\vth .
\]
(More precisely the $G_{N-2}(\vth)$ have this form for $M=1,\dots,N-2$ only. For 
$M=N-1$ the coefficient of the product of $Z_k$-s changes to 
$\frac{\vth(\vth -\frac{i}{2}(N-2))}{(\vth-\frac{i}{2}(N-1))(\vth +
\frac{i}{2}(N-2))}$ in (\ref{eq:gn2})). Setting $Z_k(\vth)\equiv 1$ for $k=1,\dots ,N-2,+,-$ corresponds to the absence of magnonic excitations, 
thus substituting these - togehter with $\mu_i^{(0)}\equiv \vth_i$ - into eq.(\ref{eq:kislam}-\ref{eq:gnnm1}) gives the eigenvalue of $\cal{D}(\vth)$ 
on the pseudovacuum. 

 
\subsection{The magnonic Bethe Ansatz equations}

The magnonic Bethe Ansatz equations are obtained from requiring the absence of poles in $\lambda$ at $\vth=\mi{i}{k}+\frac{i}{2}k$, $k=1,\dots, N-2$ and at 
$\vth=\mi{i}{\pm}+\frac{i}{2}(N-1)$. These equations have the form
\begin{multline}
 \prod_{i=1}^{n_k} \frac{\mi{j}{k}-\mi{i}{k}+i}{\mi{j}{k}-\mi{i}{k}-i} \cdot
\frac{\mi{j}{k}+\mi{i}{k}+i}{\mi{j}{k}+\mi{i}{k}-i}\times\\
 \prod_{l\in L_k} \prod_{i=1}^{n_l} \frac{\mi{j}{k}-\mi{i}{l}-\frac{i}{2}}{\mi{j}{k}-\mi{i}{l}+\frac{i}{2}} \cdot\frac{\mi{j}{k}+\mi{i}{l}-\frac{i}{2}}{\mi{j}{k}+
\mi{i}{l}+\frac{i}{2}} = - \frac{\mi{j}{k}+\frac{i}{2}}{\mi{j}{k}-\frac{i}{2}},\quad k\neq M \label{eq:knem}
\end{multline}
and
\begin{multline}
 \prod_{i=1}^{n_M} \frac{\mi{j}{M}-\mi{i}{M}+i}{\mi{j}{M}-\mi{i}{M}-i} \cdot
\frac{\mi{j}{M}+\mi{i}{M}+i}{\mi{j}{M}+\mi{i}{M}-i}\times\\
 \prod_{l\in L_M} \prod_{i=1}^{n_l} \frac{\mi{j}{M}-\mi{i}{l}-\frac{i}{2}}{\mi{j}{M}-\mi{i}{l}+\frac{i}{2}} \cdot\frac{\mi{j}{M}+\mi{i}{l}-\frac{i}{2}}{\mi{j}{M}+
\mi{i}{l}+\frac{i}{2}} =\\ 
- \frac{\mi{j}{M}+\frac{i}{2}}{\mi{j}{M}-\frac{i}{2}}
\Bigl[\frac{\mi{j}{M}-\frac{i}{2}(N-M)}{\mi{j}{M}+\frac{i}{2}(N-M)}\Bigr]^2
,\quad k=M \label{eq:km}
\end{multline}
where $L_t=\{t-1,t+1\},\{N-3,-,+\},\{N-2\}$ for $1\leq t \leq N-3$, $t=N-2$ and 
$t=\pm$ respectively. 

These equations determine $\mi{i}{k}$ $k=1,\dots, N-2,\pm$ in terms of 
$\mi{i}{0}\equiv\vth_i$. Note that substituting 
$\mi{j}{k}\rightarrow -\mi{j}{k}$ (but keeping $\mi{i}{k}$ $i\neq j$ and 
$\mi{i}{l}$ $l\neq k$ the same) in eq.(\ref{eq:knem}-\ref{eq:km}) changes both sides of the equations to their inverses, thus the $\mi{j}{k}$ roots are 
doubled, to every $\mi{j}{k}$ solving (\ref{eq:knem}-\ref{eq:km}) there is another one $-\mi{j}{k}$. 
 
Comparing these equations to the periodic ones \cite{deVega:1986xj}  
\cite{Gromov:2006dh} shows 
that they are similar to those with some differences. The differences include the $\mi{i}{k}\rightarrow -\mi{i}{k}$ doubling of all factors on the l.h.s. and the appearance of the non trivial factor on the r.h.s. One can understand the appearance of 
$- \frac{\mi{j}{k}+\frac{i}{2}}{\mi{j}{k}-\frac{i}{2}}$on the r.h.s of 
(\ref{eq:knem}-\ref{eq:km}) by realizing that this is nothing but the $i=j$ 
term of the first product on the l.h.s. Thus canceling them one gets $1$ 
(respectively $\Bigl[\frac{\mi{j}{M}-\frac{i}{2}(N-M)}{\mi{j}{M}+\frac{i}{2}(N-M)}\Bigr]^2$) on the r.h.s. of (\ref{eq:knem}) (resp. (\ref{eq:km})),  
and $\prod\limits_{i\neq j}^{n_k}$ on the l.h.s. This \lq reduced' or 
\lq simplified' form of the BAE 
may be interpreted as the Bethe-Yang equations for the magnons $k=1,\dots,N-2,+,-$
\begin{multline}
r_k^{(+)}(\mi{j}{k})r_k^{(-)}(\mi{j}{k})\prod\limits_{i\neq j}^{n_k}
S_{kk}(\mi{j}{k}-\mi{i}{k})S_{kk}(\mi{j}{k}+\mi{i}{k})\times \\
\prod\limits_{l\neq k}\prod\limits_{i=1}^{n_l}S_{kl}(\mi{j}{k}-\mi{i}{l}) S_{kl}(\mi{j}{k}+\mi{i}{l})=1, \label{eq:magBY}
\end{multline}
where $S_{kl}(\mu)$ describes the (diagonal) scattering of the $k$-th and $l$-th 
magnon on each other, while $r_k^{(+)}$  $r_k^{(-)}$ describe the 
(diagonal) reflections of the $k$-th magnon on the two ends of the strip\footnote{The absence of any $e^{2ipL}$ type term in these equations can be understood by realizing that although the magnons have rapidity they carry no momentum.}. 
The 
$S_{kl}(\mu)$ is best described by assigning the magnons to the simple roots of 
$\mathrm{D}_N$: $k=1,\dots ,N-2\rightarrow\alpha_k$, $k=+\rightarrow \alpha_{N-1}$, $k=-\rightarrow \alpha_N$ with the following non vanishing scalar products
\[
(\alpha_k\cdot\alpha_k)=2\ \ \forall k,\quad (\alpha_k\cdot\alpha_{k+1})=-1,\ k=1,\dots N-2, 
\quad(\alpha_{N-2}\cdot\alpha_N)=-1,
\]
while $\mi{i}{0}\equiv\vth_i$ to the root $\alpha_0$ satisfying $(\alpha_0\cdot\alpha_k)=-\delta_{1k}$:
\begin{equation}
S_{kl}(\mu)=\frac{\mu+\frac{i}{2}(\alpha_k\cdot\alpha_l)}{\mu-\frac{i}{2}(\alpha_k\cdot\alpha_l)}. \label{eq:magnonsmat} 
\end{equation}
This magnon-magnon $S$ matrix is of course the same as the one extracted from 
the magnonic BAE in the periodic case \cite{Gromov:2006dh}. 
The novelty is that one can extract 
also the magnonic reflections from the simplified form of 
(\ref{eq:knem}) and (\ref{eq:km}) (assuming 
they are equal on the two ends of the strip)
\[
 r_k^{(+)}(\mi{j}{k})=r_k^{(-)}(\mi{j}{k})=1,\quad k\neq M\qquad
r_M^{(+)}(\mi{j}{M})=r_M^{(-)}(\mi{j}{M})=
-\frac{\mi{j}{M}-\frac{i}{2}(N-M)}{\mi{j}{M}+\frac{i}{2}(N-M)}.
\]
  
The $N=2$ case is somewhat exceptional: in this case $\mi{i}{(N-2)}=\mi{i}{0}\equiv
\vth_i$ and there are only two functions $G_{N-2}(\vth)\equiv G_0(\vth)$ and
$G_{N-1}(\vth)\equiv G_1(\vth)$ contributing to $\lambda$ in (\ref{eq:kislam}). 
The BAE follow from the absence of poles in $\lambda$ at $\vth=\mi{i}{\pm}+\frac{i}{2}$, 
and are of the form (\ref{eq:knem}-\ref{eq:km}): 
\begin{multline}
 \prod_{i\neq j}^{n_\pm} \frac{\mi{j}{\pm}-\mi{i}{\pm}+i}{\mi{j}{\pm}-\mi{i}{\pm}-i} \cdot
\frac{\mi{j}{\pm}+\mi{i}{\pm}+i}{\mi{j}{\pm}+\mi{i}{\pm}-i}\times\\
 \prod_{i=1}^{n_0} \frac{\mi{j}{\pm}-\mi{i}{0}-\frac{i}{2}}{\mi{j}{\pm}-\mi{i}{0}+\frac{i}{2}} \cdot\frac{\mi{j}{\pm}+\mi{i}{0}-\frac{i}{2}}{\mi{j}{\pm}+
\mi{i}{0}+\frac{i}{2}} =
\begin{cases} 1 \quad &\mathrm{for}\ M=2 \\
             \Bigl(\frac{\mi{j}{\pm}-\frac{i}{2}}{\mi{j}{\pm}+\frac{i}{2}}\bigr)^2 \quad &\mathrm{for}\ M=1
\end{cases} \label{eq:o4bae}
\end{multline}
where $M=2$ corresponds to the pure Neumann, while $M=1$ to the $2$ Dirichlet and $2$ Neumann boundary conditions. 

It is well known, that the periodic $O(4)$ sigma model is equivalent to the 
$SU(2)$ principal model since its $S$ matrix factorizes into the product of 
$SU(2)$ $S$ matrices. The pure Neumann and $2$ Dirichlet $2$ Neumann boundary reflection matrices also factorize into the product of some (constant) $SU(2)$ reflection matrices \cite{Ancieto:2015ax}. Using this fact, the finite volume problem of the $O(4)$ sigma model with these boundaries can be solved in the \lq\lq $SU(2)$ language'' without any sort of nesting. This was done in 
\cite{Ancieto:2015ax} and the resulting BAE are precisely  
the ones in (\ref{eq:o4bae}) upon identifying the magnons with $k=\pm$ and the left/right $SU(2)$ magnons in the $SU(2)$ description. (This coincidence
 extends also to the eigenvalue of the transfer matrix.)    
We think this gives a strong support to our general results.

\subsection{The boundary Bethe Yang equations for the massive particles}

Finally we turn to the discussion of the boundary Bethe Yang equations 
(\ref{eq:BBYeq}) for the massive particles. For this one needs the 
$\Lambda(\vth,\{\vth_j\})$ eigenvalue of the double row transfer matrix 
$\mathcal{T}(\vth)$; to obtain this one has to multiply 
$\lambda(\vth,\{\mi{i}{k}\})$ in (\ref{eq:kislam}) 
 by the product of scalar factors 
$R_2(\vth)R_2(\hat{\vth})\prod\sigma_2(\vth-\vth_i)\sigma_2(\vth+\vth_i)$. 
It is important to notice that (\ref{eq:BBYeq}) contains 
$\Lambda(\vth_i,\{\vth_j\})$ and in the $G_0(\vth)$ function, appearing in 
$\lambda$, there is a pole at $\vth=\vth_i(\equiv\mi{i}{0})$. To cancel this 
pole we write $\sigma_2(\vth\pm\vth_i)=
\frac{\vth\pm\vth_i}{i-(\vth\pm\vth_i)}S_0(\vth\pm\vth_i)$\footnote{Note that this implies that only the term containing $G_0(\vth)$ contributes.}
with
\[
S_0(\vth)=\frac{\Gamma\Bigl(\frac{1}{2}+\frac{1}{2N-2}+\varphi\Bigr) 
			 \Gamma\Bigl(1+\varphi\Bigr)
			 \Gamma\Bigl(\frac{1}{2}-\varphi\Bigr)
			 \Gamma\Bigl(\frac{1}{2N-2}-\varphi\Bigr)}
			{\Gamma\Bigl(\frac{1}{2}+\frac{1}{2N-2}-\varphi\Bigr) 
			 \Gamma\Bigl(1-\varphi\Bigr)
			 \Gamma\Bigl(\frac{1}{2}+\varphi\Bigr)
			 \Gamma\Bigl(\frac{1}{2N-2}+\varphi\Bigr)},\quad
\varphi=\frac{i\vth}{2N-2}.
\]
Furthermore we express $R_2(\hat{\vth})$ in terms of $R_2(\vth)$ since according the boundary crossing equation (\ref{eq:Bcross}) they are related as
\[
R_2(\vth)=-\frac{\vth-i(N-1)}{\vth-\frac{i}{2}(N-1)}\cdot
\frac{\vth-\frac{i}{2}(N-2)}{\vth-\frac{i}{2}}S_0(2\vth)R_2(\hat{\vth})
\frac{\vth-\frac{i}{2}(N-2M)}{\vth-\frac{i}{2}(N+2M-2)}.
\]
This way we find from (\ref{eq:BBYeq}) the Bethe Yang equations for the particle  
rapidities in the form
\begin{equation}
e^{2ip(\vth_j)L}R^2_2(\vth_j)\Bigl(\frac{\vth_j+\frac{i}{2}(N-2M)}{\frac{i}{2}(N-2M)-\vth_j}\Bigr)^2\prod\limits_{i\neq j}^{n}S_0(\vth_j-\vth_i)S_0(\vth_j+\vth_i)
Z_1(\vth_j)
=1,\quad j=1,\dots, n,\label{eq:vegso}
\end{equation}
for $M=1,\dots, N$ and $N\geq 3$; for $N=2$ the only change is that 
$Z_1(\vth_j)$ is replaced by $Z_+(\vth_j)Z_-(\vth_j)$. The factor in front of the product term in (\ref{eq:vegso}) is nothing but $c^2(\vth_j)$, thus the $(R_2(\vth_j)c(\vth_j))^2$ term plays the role of an \lq effective' reflection factor for all the particles. This effective reflection factor reproduces the pure 
Neumann coefficient, $K(\vth_j)$, (\ref{eq:Neumann}), 
for $M=N$, since $R_2(\vth)c(\vth)\vert_{M=N}=K(\vth)$. 

Eq.(\ref{eq:vegso}) differs from the analogous periodic equation 
\cite{Gromov:2006dh} in 
addition to the appearance of this effective reflection factor also by the $\vth_i\rightarrow -\vth_i$ $\mi{i}{1}\rightarrow -\mi{i}{1}$ doubling of the terms in the various products. As a result substituting  
 $\vth_j\rightarrow -\vth_j$ on the l.h.s. of (\ref{eq:vegso}) changes 
every term to its inverse; since the $1$ on the r.h.s. is invariant, we 
conclude that if $\vth_j$ solves (\ref{eq:vegso}) then $-\vth_j$ is also a 
solution, i.e. $\vth_j$ are also doubled. 
 
\section{The case with different boundaries}

The procedure to obtain the eigenvalue of the DTM and the accompanying Bethe Ansatz equations can be used also when the the reflection matrices at the ends of the interval are different. In this chapter we summarize how the previous results change in this case. 

We assume that on the left $(+)$ and right $(-)$ ends of the interval the reflection matrices are different, but both of them are of the diagonal type 
(\ref{eq:Rsep})  
\begin{equation}
R_A^{(\pm)} (\theta)=R_2^\pm (\theta)(\mathbf{R}_a^{(\pm )} ,
\mathbf{R}_{\bar{a}}^{(\pm )} )=R_2^\pm(\theta)\mathbf{R}_{A}^{(\pm)} 
\end{equation} 
where 
\[
\mathbf{R}_{a}^{(\pm)}=\mathrm{diag}(\underbrace{c^\pm,\dots,c^\pm}_{\text M^\pm},\underbrace{1,\dots,1}_{\text N-M^\pm}),\quad c^\pm =\frac{N-2M^\pm +\theta}{N-2M^\pm -\theta},
\]
with $M^\pm$ being the two integers characterizing the left and right 
boundaries and 
\[
R_2^\pm(\theta)=R_2(\theta)\vert_{K=2M^\pm} ,
\]
with $R_2(\theta)$ in (\ref{eq:r2expl}). 

The reduced monodromy and transfer matrices are introduced now by 
\begin{equation}
\mathcal{T}(\theta)=R_2^-(\theta)R_2^+(\hth)\prod\limits_{i=1}^n\sigma_2(\theta-\theta_i)\sigma_2(\theta+\theta_i)\mathcal{D}(\theta),
\end{equation}
\begin{equation}
\omega_A(\theta,\{\theta_i\})=R_2^-(\theta)\prod\limits_{i=1}^n\sigma_2(\theta-\theta_i)\sigma_2(\theta+\theta_i)\mathbb{M}_{A}(\theta,\{\theta_i\}),
\end{equation}
where
\begin{align}
 \mathcal{D}(\theta;\{\theta_i\})&=\mathrm{tr}_A\bigl[\mathbf{R}_{A}^{(+)}(\hth)
\mathbb{M}_{A}(\theta;\{\theta_i\})\bigr],\nonumber\\
 \mathbb{M}_{A}(\theta;\{\theta_i\})&=\mathbb{T}_{A}(\theta;\{\theta_i\})
\mathbf{R}_{A}^{(-)}(\theta)\hat{\mathbb{T}}_{A}(\theta;\{\theta_i\}),\nonumber
\end{align}
with the same $\mathbb{T}$ and $\hat{\mathbb{T}}$ as in (\ref{eq:ThatT}). The equation for the eigenvalues of $\cal{D}(\theta)$ following 
from the crossing symmetry of the bulk $S$ matrix has the form now
\[
\lambda(\theta,\{\theta_i\},M^+,M^-)=\lambda(\hth,\{\theta_i\},M^-,M^+)
\]
instead of (\ref{eq:seszim}).

The $M^\pm$ integers appear in the eigenvalue of $\mathcal{D}$ in a non 
trivial way. The gross structure of $\lambda $ remains the same as in
(\ref{eq:kislam})
 \begin{align}
\lambda(\vth ,\{\mi{i}{k}\})&=
\frac{\vth -\frac{i}{2}(N-2)}{\vth-\frac{i}{2}(N-1)}\cdot
\frac{\vth -i(N-1)}{\vth-\frac{i}{2}(N+2M^+ -2)}\sum\limits_{k=0}^{N-1}G_k(\vth) 
\nonumber\\ &+
\frac{\vth -\frac{i}{2}N}{\vth-\frac{i}{2}(N-1)}\cdot
\frac{\vth}{\vth-\frac{i}{2}(N-2M^-)}\sum\limits_{k=0}^{N-1}\tilde{G}_k(\vth),
\end{align}
but the ``fine structure'' of the various $G_k(\vth)$ functions also changes. 
This means that although they retain the same product of $Z_k(\vth)$ factors as 
previously the coefficients are different now, instead of (\ref{eq:gk1}-\ref{eq:gk2}) one finds 
\begin{equation}
 G_k(\vth)= g_k^+(\vth)g_k^-(\vth)Z_k^{-1}(\vth-\frac{i}{2}(k+1))
Z_{k+1}(\vth-\frac{i}{2}k)
\end{equation}
where
\begin{align}
g_k^-(\vth)=&-\frac{\vth(\vth+\frac{i}{2}(N-2M^-))}
{(\vth-\frac{i}{2}k)(\vth-\frac{i}{2}(N-2M^-))} ,\quad k < M^-\\
g_k^-(\vth)=&\frac{\vth(\vth-\frac{i}{2}N)}
{(\vth-\frac{i}{2}k)(\vth-\frac{i}{2}(N-2M^-))} ,\quad k\geq M^-
\end{align}
and
\begin{equation}
 g_k^+(\vth)=-\frac{(\vth+\frac{i}{2}(N-2M^+))}
{(\vth-\frac{i}{2}(k+1))}, \ k < M^+ \qquad
 g_k^+(\vth)=\frac{(\vth-\frac{i}{2}N)}
{(\vth-\frac{i}{2}(k+1))},\ k\geq M^+. 
\end{equation}
($G_{N-2}(\vth)$ and $G_{N-1}(\vth)$ follow the same pattern with $g_{N-2}^\pm
(\vth)$ and $g_{N-1}^\pm (\vth)$). The relation between $G_k(\vth)$ and 
$\tilde{G}_k(\vth)$ is also slightly more complicated
\[
\tilde{G}_k(\vth)=G_k(\hat{\vth})\vert_{M^-\leftrightarrow M^+}.
\]

Compared to $\lambda$ the magnonic 
BAE and the boundary Bethe Yang equations undergo
much smaller changes as a result of the different boundaries. The simplest way 
to describe the new magnonic BAE is in terms of the effective magnonic 
Bethe Yang equations (\ref{eq:magBY}): the magnon-magnon $S$ matrix is of 
course the same as in (\ref{eq:magnonsmat}) and only the magnon reflections 
$r^{(\pm)}_k$ change. Indeed in the two boundary case the nontrivial ones are
\[
r_{M^-}^{(-)}(\mi{j}{M^-})=
-\frac{\mi{j}{M^-}-\frac{i}{2}(N-M^-)}{\mi{j}{M^-}+\frac{i}{2}(N-M^-)},\quad
r_{M^+}^{(+)}(\mi{j}{M^+})=
-\frac{\mi{j}{M^+}-\frac{i}{2}(N-M^+)}{\mi{j}{M^+}+\frac{i}{2}(N-M^+)},
\]   
all the other $r^{(\pm)}_k$ are just $1$. In a similar way the only change in 
the boundary Bethe Yang equations (\ref{eq:vegso}) is that in the two boundary 
case one must make the
\[
(R_2(\vth_j)c(\vth_j))^2\rightarrow
R_2^+(\vth_j)c^+(\vth_j)R_2^-(\vth_j)c^-(\vth_j)
\]
substitution.
   
\section{Conclusions}

In this paper we considered the $O(2N)$ sigma model with integrable diagonal boundaries and as a main result derived the boundary Bethe Yang equation for the particle rapidities (\ref{eq:vegso}) together with the boundary versions of the Bethe Ansatz equations (\ref{eq:knem}-\ref{eq:km}). We achieved this by diagonalizing the double row transfer matrix using the generalization of the method of 
\cite{deVega:1986xj}. During this investigation we paid a particular attention to the relation between the wanted and unwanted terms and 
 argued that the vanishing of the residue of the eigenvalue at the magnon rapidities is not only a necessary but also a sufficient condition. 

In this investigation we restricted our attention to the case when the particle 
reflections at the ends of the interval were 
identical and diagonal. Working in an appropriate complex basis this required that $K$, the number of fields satisfying Neumann 
b.c., had to be even, $K=2M$; and this led to the $O(2N)\rightarrow Q(2M)\times O(2(N-M))$ symmetry breaking pattern by the boundary conditions. Please note 
that in this symmetry breaking the rank of the symmetry group does not change, 
while for $K$ odd it would decresase by one. For this reason we expect that
the $K=$ odd case can not be obtained from our results by some ``analytic'' 
continuation.     

In this paper we also gave the main results for  
the case with different (diagonal) reflection matrices at the two ends of the 
interval.      

Recently considerable progress has been made to construct the eigenvalues and 
eigenvectors of DTMs constructed with non diagonal reflection matrices
\cite{Cao:2013qxa}. The boundary integrable $O(2N)$ sigma models in some cases have non diagonal reflection matrices \cite{Moriconi:2001xz}. 
It would be interesting to use the 
method of \cite{Cao:2013qxa} to solve these cases too. 

The results presented here may have some interesting applications. Working 
along the lines of \cite{Bajnok:2013sza} they may be used to establish a connection to the classical spectral curve of $O(2N)$ sigma models with integrable boundaries. In a similar way, the results presented here, may provide the basis to extend the investigation of \cite{Gromov:2006dh} - devoted to establish the relation between strings and multiparticle states of quantum sigma models - from closed strings to open ones. 

{\bf Note added} Proceeding further with the argument leading to 
eq.(\ref{eq:dkrosszim}) by exploiting also the boundary crossing property of 
the reflection matrices one can show the exact invariance 
$\mathcal{D}(\theta,\{\theta_i\})=\mathcal{D}(\hth,\{\theta_i\})$. We thank 
Rafael Nepomechie for correspondence on this issue.  

\section*{Acknowledgments}

We thank Zolt\'an Bajnok and Rafael Nepomechie for the useful discussions and for reading the manuscript. This investigation was supported in part by the 
Hungarian National Research Fund OTKA under K116505.           
\noindent 
\appendix
\renewcommand{\Am}[1]{\mathbb{A}_{#1}}
\renewcommand{\Bm}[1]{\mathbb{B}_{#1}}
\renewcommand{\Dm}[1]{\mathbb{D}_{#1}}
\renewcommand{\Dvm}[1]{\overline{\mathbb{D}}_{#1}}
\section{Notation}

We use the following notation:
\begin{enumerate}
 \item Indexes in capital letters denote $\mathbb{C}^{2N}$ vector spaces.
 \item Indexes in lower case letters denote $\mathbb{C}^{N}$ vector spaces.
 \item $\X{a}$ denotes a matrix acting on the vector space indexed by $a$. 
Alternatively we can write its indexes explicitly 
 \[
  X_{a_1}^{a_2}.
 \]
 \item $\X{ab}$ denotes a matrix acting on the tensor product of the vector spaces indexed by $a$ and $b$. Using explicit indexes we write \[X_{a_1b_1}^{a_2b_2}.\]
 \item $\Xm{a}$ is a matrix acting on the tensor product of the auxiliary vector space indexed by $a$ and the full quantum space. With explicit indexes \[\Xm{a_1}^{a_2}=X_{a_1B_1^1\dots B_1^n}^{a_2B_2^1\dots B_2^n}.\]
 \item \[\X{ab}^{t_a}\] denotes the transposition in the $a$ space.
\item In sect.3.5 $\mathbb{M}_a^{\Omega_k}$ is a $N-k+1$ dimensional matrix in 
the auxiliary space while $\mathbb{B}_a^{\Omega_k}$ ($\mathbb{C}_a^{\Omega_k}$) 
denote a $N-k$ dimensional row (respectively column) vector, and $\mathbb{D}_a^{\Omega_k}$ is a $N-k$ dimensional matrix. 
\end{enumerate}
\section{Derivation of eq.(\ref{eq4:comABDv})}

The unwanted term containing $\Am{b}$ in (\ref{eq4:CR_AB}) can be written as
 \begin{align}
 -\frac{1}{\theta-u}\operatorname*{Res}_{\theta'=u}\biggl[\Bigl\{ \Sm{ab}(u-\theta')\Bm{b}(\theta) \Qm{ab}(u+\theta) \Am{a}(\theta') \Qm{ab}^{-1}(u-\theta') \Bigr\}^{t_b} \Qm{ab}^{-1}(\hat{u}-\theta) \biggr]^{t_b}=\nonumber\\
 = -\Bm{a}(\theta)\biggl[\Bigl\{ \Qm{ab}(u+\theta) \Am{b}(u)  \Qm{ab}^{-1}(u-\theta)\Um{ab}(u-\theta) \Bigr\}^{t_b} \Qm{ab}^{-1}(\hat{u}-\theta) \biggr]^{t_b}\label{eq4:biz1}.
\end{align}
Making use of the following identities
\begin{align}
 \Qm{ab}^{-1}(u-\theta)\Um{ab}(u-\theta)&=\frac{2}{\theta-u}(\Pm{ab}-\Km{ab}),\nonumber\\
 \Pm{ab}\Am{b}\Km{ab}=\Am{b}^{t_b}\Km{ab},\qquad
 \Km{ab}\Am{b}\Pm{ab}&=\Km{ab}\Am{b}^{t_b},\qquad
 \Km{ab}\Am{b}\Km{ab}=\Km{ab}\mathrm{tr}_b\Am{b}^{t_b},\nonumber
\end{align}
the r.h.s. of eq.(\ref{eq4:biz1}) can be written in the following form
\begin{align}
 -\Bm{a}(\theta)\biggl[\Bigl\{(\Ima{ab}-\frac{2}{\hat{u}-\theta}\Km{ab})\Am{b}(u)\frac{2}{\theta-u}(\Pm{ab}-\Km{ab})\Bigr\}^{t_b}(\Ima{ab}-\frac{2}{\hat{u}-\theta+2}\Km{ab})\biggr]^{t_b}=\nonumber\\
 = - \Bm{a}(\theta)\frac{2}{\theta-u}\Bigl[\Am{b}(u)(\Pm{ab}-\Km{ab})-\frac{2}{\theta-\hat{u}}(\Pm{ab}-\Km{ab})(\mathrm{tr}_b\Am{b}(u)-\Am{b}^{t_b}(u))\Bigr]\label{eq4:biz2}.
\end{align}
The \lq unwanted' term containing $\Dm{ }$ in (\ref{eq4:CR_AB}) is written as
\begin{align}
 \frac{1}{\theta-\hat{u}}\operatorname*{Res}_{\theta'=u}\biggl[\Sm{ab}(\theta'-u)\Bm{b}(\theta) \Dm{a}(\theta')^{t_a}\Qm{ab}^{-1}(\theta'-u)\biggr]^{t_b}=\nonumber\\
 = \Bm{a}(\theta)\Bigl[\Dm{b}^{t_b}(\hat{u})\Qm{ab}^{-1}(\hat{u}-\theta)\Um{ab}(\hat{u}-\theta)\Bigr]^{t_b}= -\Bm{a}(\theta)\frac{2}{\theta-\hat{u}}(\Pm{ab}-\Km{ab})\Dm{b}(u)\label{eq4:biz3}.
\end{align}
Finally adding (\ref{eq4:biz2}) and (\ref{eq4:biz3}) we get 
\begin{align}
  - \Bm{a}(\theta)(\Pm{ab}-\Km{ab})\biggl[\frac{2}{\theta-u}\Bigl\{\Am{b}(u)-\frac{2}{u-\hat{u}}(\mathrm{tr}_b\Am{b}(u)-\Am{b}^{t_b}(u))\Bigr\}+\frac{2}{\theta-\hat{u}}\Dvm{b}(u)\biggr]=\nonumber\\ 
 =-\biggl[\Bigl\{ \Bm{a}(\theta) \Qm{ab}(2u) \Am{b}(u) \Qm{ab}^{-1}(u-\theta)\Um{ab}(u-\theta) \Bigr\}^{t_b} \Qm{ab}^{-1}(\hat{u}-u) \biggr]^{t_b}-\nonumber\\
 -\biggl[\Bm{a}(\theta) \Dvm{b}^{t_b}(u)\Qm{ab}^{-1}(\hat{u}-\theta)\Um{ab}(\hat{u}-\theta)\biggr]^{t_b},
\end{align}
which is equivalent to the unwanted terms in (\ref{eq4:comABDv}).

\renewcommand{\An}[2]{\mathbb{A}_{#1}^{\Omega_#2}}
\newcommand{\At}[2]{\mathbb{A}_{#1}^{\Omega_#2t_{#1}}}
\renewcommand{\Bm}[2]{\mathbb{B}_{#1}^{\Omega_#2}}
\newcommand{\Bt}[2]{\mathbb{B}_{#1}^{\Omega_#2t_{#1}}}
\renewcommand{\Dm}[2]{\mathbb{D}_{#1}^{\Omega_#2}}
\renewcommand{\Dvm}[2]{\overline{\mathbb{D}}_{#1}^{\Omega_#2}}
\renewcommand{\Dhm}[2]{\widetilde{\mathbb{D}}_{#1}^{\Omega_#2}}
\newcommand{\Dvt}[2]{\overline{\mathbb{D}}_{#1}^{\Omega_#2t_{#1}}}
\newcommand{\Dht}[2]{\widetilde{\mathbb{D}}_{#1}^{\Omega_#2t_{#1}}}

\section{The relation between the wanted and unwanted terms}

In this appendix we study the relation between the wanted and unwanted terms 
appearing in the commutator of the transfer matrix $\mathcal{D}(\theta)$, 
eq.(\ref{eq:DTMr2}), and $\mathbb{B}^{ij}(u)$. 

First we rewrite 
$\mathcal{D}(\theta)\equiv
\D{0}(\theta)$\footnote{To emphasize that the $\mathbb{A}$, $\dots$, 
$\mathbb{D}$ operators act on the space $\Omega^{0}$ and to distinguish them from the various similar operators appearing in the following sections we denote them by an upper index $\Omega_0$}  in a more symmetric form; to this end we define $\Dhm{a}{0}(\theta)$ by the relation
\begin{equation}
 \Dvm{a}{0}(\theta) = \Dhm{a}{0}(\theta) + \frac{2}{\theta-\hth} \left( \Dht{a}{0}(\theta) - \mathrm{tr}_a \Dhm{a}{0}(\theta) \right),
\end{equation}
since using it $\D{0}(\theta)$ can be written as
\begin{equation}
 \D0(\theta) = \frac{\theta-N+2}{\theta-N+1}\cdot\frac{\theta-2N+2}{\theta-N-2M+2}\Bigl(
\mathrm{tr}_a \left[ \Rvp{a}{1}(\hth) \An{a}{0}(\theta) \right] + \mathrm{tr}_a \left[ \Rvp{a}{1}(\hth) \Dhm{a}{0}(\theta) \right]\Bigr) ,
\end{equation}
with $\Rvp{a}{1}$ being the constant diagonal matrix introduced in 
(\ref{eq:R1bar})
\[
 \Rvp{a}{1} = \begin{pmatrix}
-\mathbf{I}_M & \\
& \mathbf{I}_{N-M} \end{pmatrix} .
\]
After some effort one can show, that the commutation relations of sect.3.3. 
imply
\newcommand{\tr}[1]{\mathrm{tr}_{#1}}
\newcommand{\wA}{\mathrm{wantA}}
\newcommand{\wD}{\mathrm{wantD}}
\newcommand{\uwA}[1]{\mathrm{uwantA#1}}
\newcommand{\uwD}[1]{\mathrm{uwantD#1}}
\begin{align}
 \mathrm{tr}_a \left[ \Rvp{a}{1} \An{a}{0}(\theta) \right] \Bm{b}{0}(u) &= \wA + \uwA1 + \uwA2,\\
 \mathrm{tr}_a \left[ \Rvp{a}{1} \Dhm{a}{0}(\theta) \right] \Bm{b}{0}(u) &= \wD + \uwD1 + \uwD2,
\end{align}
where $\wA$ and $\wD$ denote the wanted terms
\begin{align}
 \wA &= \Bm{ij}{0}(u) \tr{a} \left[ \Rvp{a}{1} \Sm{ai}^{t_i}(u-\theta) \Sm{aj}^{t_j}(\hat{u}-\theta) \An{a}{0}(\theta) \Qm{aj}^{-1}(u-\theta) \Qm{ai}^{-1}(\hat{u}-\theta) \right],\\ 
 \wD &= \Bm{ij}{0}(u) \tr{a} \left[ \Rvp{a}{1} \Qm{ai}^{-1t_i}(\theta-u) \Qm{aj}^{-1t_j}(\theta-\hat{u}) \Dhm{a}{0}(\theta) \Sm{aj}(\theta-u) \Sm{ai}(\theta-\hat{u}) \right], 
\end{align}
(note, that using some simple identities between $\Sm{ab}$ and $\Qm{ab}$ we recast these in a more suitable form), while the various unwanted terms have the 
form
\begin{align}
 \mathrm{\uwA1} =& \frac{2}{\theta-u} \left\{ \left[ \left( \Rvp{b}{1} \Bm{b}{0}(\theta) - \Bt{b}{0}(\theta) \Rvp{b}{1} \right) \At{b}{0}(u) \right]^{t_b} -\right. \nonumber\\
 & \left. \frac{1}{N-u-1} \left( \Rvp{b}{1} \Bm{b}{0}(\theta) - \Bt{b}{0}(\theta) \Rvp{b}{1} \right) \left( \At{b}{0}(u) - \mathrm{tr}_b \At{b}{0}(u) \right) \right\} \\
 \mathrm{\uwA2} =& \frac{2}{\theta-\hat{u}} \left( \Rvp{b}{1} \Bm{b}{0}(\theta) - \Bt{b}{0}(\theta) \Rvp{b}{1} \right) \Dvm{b}{0}(u),\\
 \mathrm{\uwD1} =& -\frac{2}{\theta-u} \frac{1}{N-2-\theta} \left[ \Rvp{b}{1} \left( \Bm{b}{0}(\theta) + (N-\theta-1) \Bt{b}{0}(\theta) \right) - \right. \nonumber\\
 & \left.  \left( \Bm{b}{0}(\theta) (N-\theta-1) + \Bt{b}{0}(\theta) \right) \Rvp{b}{1} \right] \Dvm{b}{0}(u),\\
 \mathrm{\uwD2} =& -\frac{2}{\theta-\hat{u}} \frac{1}{N-2-\theta} \left\{ \left[ \left( \Rvp{b}{1} \left( \Bm{b}{0}(\theta) + (N-\theta-1) \Bt{b}{0}(\theta) \right) - \right. \right. \right. \nonumber\\
 & \left.  \left.  \left( \Bm{b}{0}(\theta) (N-\theta-1) + \Bt{b}{0}(\theta) \right) \Rvp{b}{1} \right) \At{b}{0}(u) \right]^{t_b} - \nonumber\\
 & \frac{1}{N-1-u} \left[ \Rvp{b}{1} \left( \Bm{b}{0}(\theta) + (N-\theta-1) \Bt{b}{0}(\theta) \right) - \right. \nonumber\\
 & \left.  \left.  \left( \Bm{b}{0}(\theta) (N-\theta-1) + \Bt{b}{0}(\theta) \right) \Rvp{b}{1} \right] \left( \At{b}{0}(u) - \mathrm{tr}_b \At{b}{0}(u) \right) \right\}.
\end{align}
Note that on the r.h.s of these equations we kept $\Dvm{b}{0}$, the reason being that it has nicer properties than $\Dhm{b}{0}$, e.g. it has a simpler action 
on the pseudovacuum. 

In these expressions $\Bm{ij}{0}$ may be viewed as a row vector in both spaces indexed by $i$ and $j$. Taking the residues of the wanted terms gives
  \begin{align}
 \Res{\theta}{u} \wA &= - \Bm{ij}{0}(u) \Rvp{i}{1} \left( \Ima{ij} - \Pm{ij} \right) \left[ -\At{i}{0}(u) - \frac{1}{N-u-1} \left( \At{j}{0}(u) - \tr{j} \An{j}{0}(u) \right) \right],\\
 \Res{\theta}{u} \wD &= \Bm{ij}{0}(u) \frac{1}{N-2-u} \left[ \Rvp{i}{1} - (N-u-1) \Rvp{j}{1} \right]  \left( \Ima{ij} - \Pm{ij} \right) \Dvm{j}{0}(u).
\end{align}
Using explicit indexes one can write the unwanted terms as
 \begin{align}
 \uwA1 =& \frac{1}{\theta-u} \Bm{ij}{0}(u) \Rvp{i}{1} \left( \Ima{ij} - \Pm{ij} \right) \left[ -\At{i}{0}(u) - \frac{1}{N-u-1} \left( \At{j}{0}(u) - \tr{j} \An{j}{0}(u) \right) \right],\\
 \uwA2 =& \frac{1}{\theta-\hat{u}} \Bm{ij}{0}(u) \Rvp{i}{1} \left( \Ima{ij} - \Pm{ij} \right) \Dvm{j}{0}(u), \\
 \uwD1 =& - \frac{1}{\theta-u} \Bm{ij}{0}(u) \frac{1}{N-2-\theta} \left[ \Rvp{i}{1} - (N-\theta-1) \Rvp{j}{1} \right]  \left( \Ima{ij} - \Pm{ij} \right) \Dvm{j}{0}(u), \\
 \uwD2 =& - \frac{1}{\theta-\hat{u}} \Bm{ij}{0}(u) \frac{1}{N-2-\theta} \left[ \Rvp{i}{1} - (N-\theta-1) \Rvp{j}{1} \right]  \left( \Ima{ij} - \Pm{ij} \right) \times \nonumber\\
 & \left[ -\At{i}{0}(u) - \frac{1}{N-u-1} \left( \At{j}{0}(u) - \tr{j} \An{j}{0}(u) \right) \right] ,
\end{align}
showing a strikingly similar structure to the residue of the wanted terms. To make this relation precise we introduce $\Res{\theta}{u} \left[ \mathrm{wanted} \right]_{ij}$ and $\left[ \mathrm{uwanted} \right]_{ij}$ by the relations
\begin{align}
 \Bm{ij}{0}(u) \Res{\theta}{u} \left[ \mathrm{wanted} \right]_{ij} &= \wA + \wD,\\
 \Bm{ij}{0}(\theta) \left[ \mathrm{uwanted} \right]_{ij} &=	\uwA1 + \uwA2 + \uwD1 + \uwD2.
\end{align}
Then, by direct computations, one can verify that 
\begin{equation}
 \left[ \mathrm{unwanted} \right]_{ij} = - \left( \frac{1}{\theta-u} + \frac{1}{\theta-\hat{u}} \right) \widetilde{\mathbf{R}}_{ij} \left[ \Res{\theta}{u} \mathrm{wanted} \right]_{ij},\label{eq:residue}
\end{equation} 
where
\begin{equation}
 \widetilde{\mathbf{R}}_{ij} = \begin{pmatrix}
 \mathbf{I}_{M^2} & & & \\
 & r\mathbf{I}_{M(N-M)} & & \\
 & & r\mathbf{I}_{M(N-M)} & \\
 & & & \mathbf{I}_{(N-M)^2} \end{pmatrix},\qquad r = \frac{N-u-2}{N-\theta-2}.
\end{equation}
In subsections 3.4-3.5 the eigenvectors of the wanted terms are constructed. 
When we let both sides of (\ref{eq:residue}) to act on these eigenvectors we 
see that demanding the vanishing of the residue of the eigenvalue is equivalent to the vanishing of the contributions of the unwanted terms. 

In the $k$-th step of the nesting procedure one can obtain in a similar way 
the sufficient condition for the vanishing of the unwanted terms in the 
commutator of $\D{k}(\theta)$, (\ref{eq:kikD}), and $\Bm{i}{k}(u)$. Indeed 
defining now $\Res{\theta}{u} \left[ \mathrm{wanted}^k \right]_{i}$ and 
$\left[ \mathrm{uwanted}^k \right]_{i}$ by the relations 
\begin{align}
 \Bm{i}{k}(u) \Res{\theta}{u} \left[ \mathrm{wanted}^k \right]_{i} &= \wA^k + \wD^k,\\
 \Bm{i}{k}(\theta) \left[ \mathrm{uwanted}^k \right]_{i} &=	\uwA1^k + \uwA2^k + \uwD1^k + \uwD2^k,
\end{align}
(where the r.h.s. of these equations denote the wanted/respectively unwanted 
terms of the commutator) after a direct computation one finds the following relation
\begin{equation}
 \left[ \mathrm{unwanted}^k \right]_{i} = - \left( \frac{1}{\theta-u} + \frac{1}{\theta+u-2k} \right) \widetilde{\mathbf{R}}_{i}^k \left[ \Res{\theta}{u} \mathrm{wanted}^k \right]_{i},\label{eq:kikresidue}
\end{equation}
with  
\begin{align}
 \widetilde{\mathbf{R}}_{i}^k &= \begin{pmatrix}
 \mathbf{I}_{M-k} & \\
 & r\mathbf{I}_{N-M} \end{pmatrix}, & k&=1,\dots,M\\
 \widetilde{\mathbf{R}}_{i}^k &= \begin{pmatrix}
 \mathbf{I}_{N-k} \end{pmatrix}, & k&=M+1,\dots,N-1\\
\end{align}
and 
\begin{equation}
 r = \frac{N-u-2M}{N-\theta-2M}.
\end{equation}
Letting both sides of eq.(\ref{eq:kikresidue}) to act on $\vert\Psi^{\Omega_k}\rangle$ one can see that (\ref{eq:kikLam}) is indeed a sufficient condition.

\bibliographystyle{JHEP}
\bibliography{intads_rev,refs}

\end{document}